\documentclass[fleqn,usenatbib]{mnras}

\usepackage{newtxtext,newtxmath}

\usepackage[T1]{fontenc}
\usepackage{ae,aecompl}

\usepackage{graphicx}	
\usepackage{amsmath}	

\usepackage{amssymb}	
\usepackage[dvipsnames,usenames]{color} 

\usepackage[table]{xcolor}

\usepackage{lastpage}
\usepackage{natbib}
\usepackage{setspace}
\usepackage[utf8]{inputenc}

\newcommand{\rsun}{{R$_\odot$} } 

\title[Central hydrogen abundance of the 16 Cyg A and B ]{On the stellar core physics of the 16 Cyg binary system: constraining the central hydrogen abundance using asteroseismology}

\author[Nsamba et al.]{
Benard Nsamba$^{1,2,4}$\thanks{E-mail: nsamba@mpa-garching.mpg.de},
Margarida S. Cunha$^{2}$\thanks{E-mail: mcunha@astro.up.pt},
Catarina I. S. A. Rocha$^{2}$, 
Cristiano J. G. N. Pereira$^{2}$
\newauthor
M\'{a}rio J. P. F. G. Monteiro$^{2,3}$, 
and
Tiago L. Campante$^{2,3}$
\\
$^{1}$ Max-Planck-Institut f\"{u}r Astrophysik, Karl-Schwarzschild-Str. 1, D-85748 Garching, Germany\\
$^{2}$Instituto de Astrof\'{\i}sica e Ci\^{e}ncias do Espa\c{c}o, Universidade do Porto, CAUP, Rua das Estrelas, PT4150-762 Porto, Portugal\\
$^{3}$Departamento de F\'{\i}sica e Astronomia, Faculdade de Ci\^{e}ncias da Universidade do Porto, Rua do Campo Alegre, s/n, PT4169-007\\ Porto, Portugal \\
$^4$Kyambogo University, Faculty of Science, Department of Physics, P.O. Box 1, Kyambogo, Kampala - Uganda
}

\date{Accepted 2022 May 7. Received 2022 May 7; in original form 2022 January 11}

\pubyear{2022}

\begin{document}
\label{firstpage}
\pagerange{\pageref{firstpage}--\pageref{lastpage}}
\maketitle

\begin{abstract}
The unprecedented quality of the asteroseismic data of solar-type stars made available by space missions such as  NASA's {\it{Kepler}} telescope are making it possible to explore stellar interior structures. This offers possibilities of constraining stellar core properties (such as core sizes, abundances, and physics) paving the way for improving the precision of the inferred stellar ages. We employ 16 Cyg A and B as our benchmark stars for an asteroseismic study in which we present a novel approach aimed at selecting from a sample of acceptable stellar models returned from Forward Modelling techniques, down to the ones that better represent the core of each star. This is accomplished by comparing specific properties of the observed frequency ratios for each star to the ones derived from the acceptable stellar models. We demonstrate that in this way we are able to constrain further the hydrogen mass fraction in the core, establishing the stars' precise evolutionary states and ages. The ranges of the derived core hydrogen mass fractions are [0.01 -- 0.06] and [0.12 -- 0.19] for 16 Cyg A and B, respectively, and, considering that the stars are coeval, the age and metal mass fraction parameters span the region [6.4 -- 7.4] Gyr and [0.023 -- 0.026], respectively.
In addition, our findings show that using a single helium-to-heavy element enrichment ratio, ($\Delta Y/\Delta Z$), when forward modelling the 16 Cyg binary system, may result in a sample of acceptable models that do not simultaneously fit the observed frequency ratios, further highlighting that such an approach to the definition of the helium content of the star may not be adequate in studies of individual stars.

\end{abstract}


\begin{keywords}
asteroseismology--stars: evolution--stars: composition--stars: oscillations--methods: statistical--stars: fundamental parameters--stars: abundances
\end{keywords}

\section{Introduction}
\label{Intro}
Photometric space missions such as the French-led CoRoT (Convection, Rotation and planetary Transits; \citealt{2008Mich,2009Baglin}) satellite, NASA's {\it{Kepler}} space telescope \citep{2010Borucki,2010Koch}, NASA's {\it{TESS}} (Transiting Exoplanet Survey Satellite; \citealt{2015Ricker}) and the future PLATO (PLAnetary Transits and Oscillations of stars; \citealt{2014Rauer}) mission are at the forefront of yielding high quality asteroseismic data of solar-type stars. Combining the seismic data with stellar atmospheric constraints, such as spectroscopic parameters (i.e. metallicity, [Fe/H], and effective temperature, $T_{\rm{eff}}$), and interferometric measurements (i.e. angular diameter), sets a platform for constraining stellar interior physics and fundamental parameters to unprecedented levels (e.g. \citealt{Metcalfe_2010,Metcalfe_2012,2014Pinsonneault,2015Valle,2017Aguirre,2018Joyce,2019Campante,2019Bellinger,2019Andreas,2020Ball,2020Jiang,2020Bowman,2020Farnir,2021Nsamba,2021Deal}, among others). In addition, this information has been employed in the precise characterisation of exoplanetary systems (e.g. \citealt{2015Campante,2020Toledo, 2020Mortier}). 

To explore the valuable seismic data made available from space missions, various  tools have been and continue to be developed or modified on the modelling side. 
These range from stellar modelling tools employed to explore stellar physics (e.g. \citealt{2008Demarque,2008Christensen,2008Weiss,2015Paxton,2018Paxton}), optimisation tools which make use of either Bayesian techniques, Markov chain Monte Carlo (MCMC) methods, or Machine learning algorithms aimed at examining the statistical relationships between stellar models and observational data (e.g. \citealt{2015Silva,2016Bellinger,2017Angelou,2019Rendle,2021aLyttle,2021Chen,Remple2021}), 
and tools aimed at exploring different regions in the stellar interior structure such as acoustic glitch analysis methods. Acoustic glitches are localized sharp variations in the sound speed caused by the ionization zones, as well as by sharp variations in the thermal stratification or in the mean molecular weight at the transition between radiative and convective regions. Acoustic glitches impose specific signatures on the stellar oscillation frequencies, that may be used as a diagnostics of the helium ionization zone, helium abundance in the envelope, and position of the base of the envelope convection zone or edge of the convective core \citep{1988Vorontsov,1990Gough,1998Thompson,2000Monteiro,2004Basu,2007Houdek,2014Mazumdar,2014Verma,2020Avel}.

Unlike in the Sun, only oscillation frequencies of modes of degree $l \leqslant 3$ are observed in solar-type stars. Consequently, over the years studies concerning the inference of detailed information on the stars' internal structure have focused on modes of low degree. Some of these studies have aimed at finding ways to
probe the stellar core sizes and examine the physics and physical processes taking place at the core edge (e.g. \citealt{2005Provost,2007Cunha,2010A&ADeheuvels,2011Cunha,2013Silva,2014Brand}).
However, modes of  $l = 3$ have been observed in only several tens of cases (e.g. \citealt{2010Bruntt,Metcalfe_2010,2017Lund}). Therefore, insights into the deep stellar interior (especially the core structure and size) are in most cases carried out through a combination of oscillation frequencies of spherical degree $l = 0, 1$ and to a less extent $l = 0, 2$ (e.g. \citealt{2005Popielski,2011Aguirre,2011Cunha,2020Rocha}).

In order to isolate information on the stellar interior, the oscillation frequencies of these low degree acoustic modes should be combined in a way such that the combination retains information about the stellar interior structure and is, simultaneously, mostly independent of the structure of the stellar outer layers. \citet{2003Roxburgh} demonstrated that this can be obtained through the computation of the ratios of the small to large frequency separations. These are constructed using five-point separations defined as (\citealt{2005Roxburgh,2013Roxburgh}):
\begin{equation}
d_{01}(n) = \frac{1}{8} (\nu_{n-1,0} - 4\nu_{n-1,1} + 6\nu_{n,0} - 4\nu_{n,1} + \nu_{n+1,0})~,
    \label{eq1}
\end{equation}
\begin{equation}
d_{10}(n) = -\frac{1}{8} (\nu_{n-1,1} - 4\nu_{n,0} + 6\nu_{n,1} - 4\nu_{n+1,0} + \nu_{n+1,1})~.
    \label{eq2}
\end{equation}
The ratios of small to large frequency separations are then defined as:
\begin{equation}
    r_{01}(n) = \frac{d_{01}(n)}{\Delta\nu_{1}(n)}~,
    \label{eq3}
\end{equation}
\begin{equation}
    r_{10}(n) = \frac{d_{10}(n)}{\Delta\nu_{0}(n+1)}~,
    \label{eq4}
\end{equation}
where $0$ and $1$ represent the radial and dipole mode degrees, respectively, $\Delta\nu_{l}$ is the the separation between modes of the same spherical degree, $l$, and consecutive radial order, $n$, expressed as $\Delta\nu_{l}~=~\nu_{l,n} - \nu_{l, n-1}$ \citep{1986Ulrich}.

The ratios $r_{01}(n)$ and $r_{10}(n)$ are also often combined as
\begin{equation}
\begin{aligned}
    r_{010} =& \{r_{01}(n),r_{10}(n),r_{01}(n+1),r_{10}(n+1),r_{01}(n+2),\\
             & r_{10}(n+2), ...\}~.
\end{aligned}
\end{equation}

These ratios provide a diagnostic of the stellar interior alone and are almost not affected by the so called ``near-surface effects'' (see \citealt{1988Christensen,1988Dziembowski,1997Christensen}). The ratios of degree $l = 0$ and $l = 1$ shown in Eq.~(\ref{eq3}) and Eq.~(\ref{eq4}) have been used to determine the base of the convective envelope in the Sun and other solar-type stars (e.g. \citealt{2009Roxburgh,2012Lebreton,2012Mazumdar,2013Silva}). In addition, during the examination of the efficiency of ratios $r_{010}$ as a seismic indicator of the presence and size of a convective core, \citet{2016Deheuvels} demonstrated that the trend of the ratios $r_{010}$ for stellar models of main-sequence stars can be approximated by quadratic polynomials. Following a similar seismic diagnostic procedure, \citet{2020Viani} showed that one of the coefficients, determined by fitting the second-order polynomial to ratios $r_{010}$ as suggested by \citet{2016Deheuvels}, can be used as an indicator of the amount of core overshoot needed to model a particular star. Furthermore, \citet{2020Viani} were also able to quantify the amount of core overshoot based on ratios $r_{010}$ for a set of {\it{Kepler}} Legacy stars and highlighted hints of a possible trend between stellar mass and core overshoot. Earlier works by \citet{2010Brand} and \citet{2011Aguirre} also argued that the ratios $r_{010}$ are not only sensitive to the presence and size of a stellar core but they are also affected by the central hydrogen content, thus can be used as indicators of the evolutionary state of a star.  

Taking 16 Cyg A and B as our benchmark stars, 
we present a novel method aimed at reducing the number of stellar models accepted by the Forward Modelling Technique (involving fitting observed oscillation frequencies and a set of atmospheric constraints) down to the ones that better represent the core of each star. This is attained by a characterisation and comparison of the ratios computed from the models and observations, following an approach similar to that proposed by \citet{2016Deheuvels}. We  demonstrate that we are able to constrain further the fraction of hydrogen in the core of both of our benchmark stars, establishing their precise evolutionary state, and stellar ages.  

This article is organized as follows. In Section~\ref{observables}, we describe our sample which is composed of two stars, their corresponding sets of observations, the details of the stellar grids, the optimization routines, and the frequency ratio fitting procedures. In Section~\ref{results}, we present our results and discussions, while in Section~\ref{conclusions} we conclude.

\section{Constraints, Models, and fitting process}
\label{observables}
The seismic and atmospheric constraints used in the optimisation process are described in Section~\ref{constraints}. Section~\ref{models} provides
a description of the stellar grids and model selection process, while Section~\ref{ratios} details how the ratios $r_{010}$ are used to add extra constraints on the central hydrogen abundance, starting from the accepted models obtained from the forward modelling process.
\subsection{Observational and known properties}
\label{constraints}
We consider the well-studied solar analogues 16 Cyg A and B as our benchmark stars. These stars are in a binary system with precisely measured angular diameters, thus their interferometric radii are available. These constraints are readily available from \citet{2013White}, who made use of the PAVO (Precision Astronomical Visual Observations; \citealt{2008Ireland}) beam combiner at the CHARA (Center for High Angular Resolution Astronomy; \citealt{ten_Brummelaar_2005}) Array. Specifically, the authors derived linear radii $R_{\rm A} = 1.22 \pm 0.02$~\rsun  and $R_{\rm B} = 1.12 \pm 0.02$~\rsun for 16 Cyg A and B, respectively. 

16 Cyg A and B are among the brightest solar-type stars observed continuously for approximately 2.5 years by the {\it{Kepler}} space telescope, yielding oscillations with exceptional signal-to-noise, allowing for detailed asteroseismic studies (e.g. \citealt{2015Metcalfe,2016Bellinger,2020Farnir}).  
The seismic data for both these stars have been analysed by \cite{2017Lund}, who extracted frequencies for more than 48 oscillation modes. 
Through the analysis of the acoustic glitch signature on the oscillation frequencies arising from the helium ionization zone, \citet{2014Verma}  constrained the surface helium mass fractions of 16 Cyg A and B to be within the intervals $Y_{\rm surf, A}~\in~[0.231, 0.251]$ and $Y_{\rm{surf, B}}~\in~[0.218, 0.266]$, respectively.
Finally, the spectroscopic parameters of 16 Cyg A and B adopted here are from  \citet{2009Ram}, specifically, $T_{\rm eff, A}~=~5825 \pm 50$~K, [Fe/H]$_{\rm A}~=~0.10 \pm 0.03$~(dex), $T_{\rm eff, B}~=~5750 \pm 50$~K, and [Fe/H]$_{\rm B}~=~0.05 \pm 0.02$~(dex). 
We stress that the observational constraints adopted in our optimisation process described in Section~\ref{models} are $T_{\rm eff}$, [Fe/H], and individual oscillation frequencies only. 

\subsection{Asteroseismic modelling and optimisation process}
\label{models}
We adopted the three stellar model grids (namely, G$_{1.4}$, G$_{2.0}$, and G$_{\rm free}$) from \citet{2021Nsamba}\footnote{Note the change in nomenclature
of the grids. Grid G$_{1.4}$, G$_{2.0}$, and G$_{\rm free}$ correspond to grid B, C, and A, respectively, in \citet{2021Nsamba}.} which were constructed using a 1D stellar evolution code (MESA\footnote{Modules for Experiments in Stellar Astrophysics} version 9793; \citealt{2015Paxton,2018Paxton}). All these grids have the same physics inputs, differing only in the treatment of the initial helium mass fraction. 
\begin{table}
\centering
\caption{Stellar grid variations}
\label{grids}
\begin{tabular}{c c  }        
\hline 
Grid name   & $\Delta Y / \Delta Z$\\
\hline \hline
G$_{1.4}$  &     1.4     \\
G$_{2.0}$  &     2.0   \\
G$_{\rm free}$  &   None   \\

\hline                                 
\end{tabular}
\end{table}
In grids G$_{1.4}$ and G$_{2.0}$, the initial helium mass fraction, $Y_{i}$, is estimated via a helium-to-heavy element enrichment ratio, ($\Delta Y/\Delta Z$), using the expression
\begin{equation}
    Y_i = \left(\frac{\Delta Y}{\Delta Z} \right) Z_i + Y_0 ,
    \label{law}
\end{equation}
where $Z_i$ is the initial metal mass fraction and $Y_0$ is the primordial big bang nucleosynthesis helium mass fraction value taken as 0.2484 \citep{2003Cyburt}. Table~\ref{grids} highlights the  helium-to-heavy element enrichment ratio used in each grid. A brief highlight of the grid dimensions is given below (refer to \citealt{2021Nsamba} for a detailed description of the uniform model physics used in all the grids);
\begin{itemize}
    \item [-] $M$ $\in$ [0.7 -- 1.25] M$_\odot$ in steps of 0.05 M$_\odot$;
    \item [-] $Z_i$ $\in$ [0.004 -- 0.04] in steps of 0.002;
    \item [-] $\alpha_{\rm{mlt}}$  $\in$ [1.2 -- 3.0] in steps of 0.2.
\end{itemize}
In grid G$_{\rm free}$, the initial helium mass fraction, $Y_i$, is in the range [0.22 -- 0.32] in steps of 0.02. The stellar models in all the grids range from the zero-age main sequence (ZAMS) to the end of the main-sequence phase, i.e. terminal-age main sequence (TAMS). For each of the models, their corresponding adiabatic oscillation frequencies for the spherical mode degrees $l = 0, 1, 2,$ and $3$ were calculated using GYRE \citep{2013Townsend}.

In order to select models with observables comparable to the observations (commonly referred to as best-fit models, acceptable models, or optimal models) and derive the stellar properties of our target stars, we make use of the grid-based optimisation tool AIMS (Asteroseismic Inference on a Massive Scale; \citealt{2016Reese,2019Rendle}). For additional details on the AIMS code, please refer to the AIMS documentation\footnote{https://gitlab.com/sasp/aims}.
In a nutshell, AIMS combines MCMC (Markov Chain Monte Carlo) and Bayesian schemes to generate a sample of representative stellar models that fit a specific set of classical and seismic constraints of a given star. For both components of our binary, the individual oscillation frequencies were used as seismic constraints while the effective temperature, $T_{\rm{eff}}$, and  metallicity, [Fe/H], were the specified classical constraints (see Section~\ref{constraints}).

It is worth noting that a well known offset hindering a direct comparison between the model frequencies and observed oscillation frequencies exists which needs to be corrected for (i.e. surface effects; \citealt{1988Christensen,1988Dziembowski,1997Christensen}). We used the two-term surface correction empirical formula suggested by \citet{2014Ball} to rectify the offsets between the model and observed frequencies. This empirical expression describes the frequency offset ($\delta \nu$) as a combination of the cubic and inverse term, and takes the form
\begin{equation}
    \delta \nu = I^{-1}(Af^{-1} + Bf^{3})~~, 
    \label{correction}
\end{equation}
where A and B are free parameters, $I$ is the mode inertia, and $f = \nu / \nu_{\rm{ac}}$. Here, $\nu$ is the oscillation frequency and $\nu_{\rm{ac}}$ is the acoustic cut-off frequency which is linearly related to the frequency of maximum power, $\nu_{\rm{max}}$ (\citealt{1991Brown,1995Kjeldsen}).

In the selection of best-fitting models, we consider a total $\chi^2$ which is a combination of seismic and classical constraints. 
We highlight that a 3$\sigma$ uncertainty cutoff on the classical constraints was applied and the best-fit models are samples of the multivariate likelihood distribution bounded solely by the criteria of $3\sigma$ on the non seismic data.
Furthermore, AIMS allows for a choice of different weights to be applied to the classical and seismic constraints. 
In this work we chose these weights such that the $\chi_{\rm{total}}^2$ function to be used in the definition of the likelihood is expressed as
\begin{equation}
    \chi_{\rm{total}}^2 = \frac{N_{c}}{N_{\nu}} \left(\chi_{\rm{seismic}}^2\right) + \chi_{\rm{classical}}^2~,
\end{equation}
where $N_{c}/N_{\nu}$ is the ratio of the number of classical constraints to the number of seismic constraints, $\chi_{\rm{seismic}}^2$ and $\chi_{\rm{classical}}^2$ are defined as sums of terms $\chi_{i}^2$ with the form
\begin{equation}
    \chi_{i}^2 = \left( \frac{O_i - \theta_i}{\sigma_i}  \right)^2 ~,
\end{equation}
where $O_i$, $\theta_i$, and $\sigma_i$ are the observed value, the model value, and the observational uncertainty, respectively. The observable constraints adopted are seismic (i.e. individual oscillation frequencies) and classical constraints (i.e. $T_{\rm eff}$ and [Fe/H]).

The treatment of weights given to the classical and seismic constraints in the model selection process is currently a subject of interest to stellar modellers and has been extensively addressed in the ``PLATO hare and hounds'' exercise for modelling main-sequence stars \citep{2021Cunha}. Finally, the $\chi_{\rm{total}}^2$ is used to obtain the likelihood function, from which the posterior probability
distributions (PDF) for the different stellar properties and their uncertainties are calculated, i.e., inform of the statistical mean and standard deviation, respectively. 

\subsection{Seismic probe of the core: fitting frequency ratios}
\label{ratios}

The inability to adequately model the stellar surface effects on the pulsation frequencies has led authors to seek frequency combinations or fitting procedures that are insensitive to those surface layers \citep{2003Roxburgh,2007Cunha,2016roxburgh} and to explore their potential for revealing the physical conditions near the stellar core \citep{2011Cunha,2011Aguirre,2014brandao}. In particular, these studies have shown that the slope ({i.e., the frequency derivative)} of the ratios of small to large frequency separations is sensitive to the gradient of the sound speed in the core, holding information on the stellar age (specifically, on the central hydrogen abundance), as well as on the size of the chemical discontinuity built at the edge of convective cores and on the amount of core overshoot. Following on these works, \cite{2016Deheuvels} argued that the combination of parameters $(a_0,a_1)$ resulting from fitting a second order polynomial of the form
\begin{equation}
    P(\nu) = a_0 + a_1\left({\nu} -\beta  \right) + a_2\left({\nu} - \gamma_1 \right)\left({\nu} - \gamma_2   \right)~
    \label{poly_fits_sd}
\end{equation}
to the ratios $r_{010}$, is particularly useful for establishing the evolutionary state of the star and the presence and size of stellar convective cores. 
\begin{figure*}
	\includegraphics[width=\columnwidth]{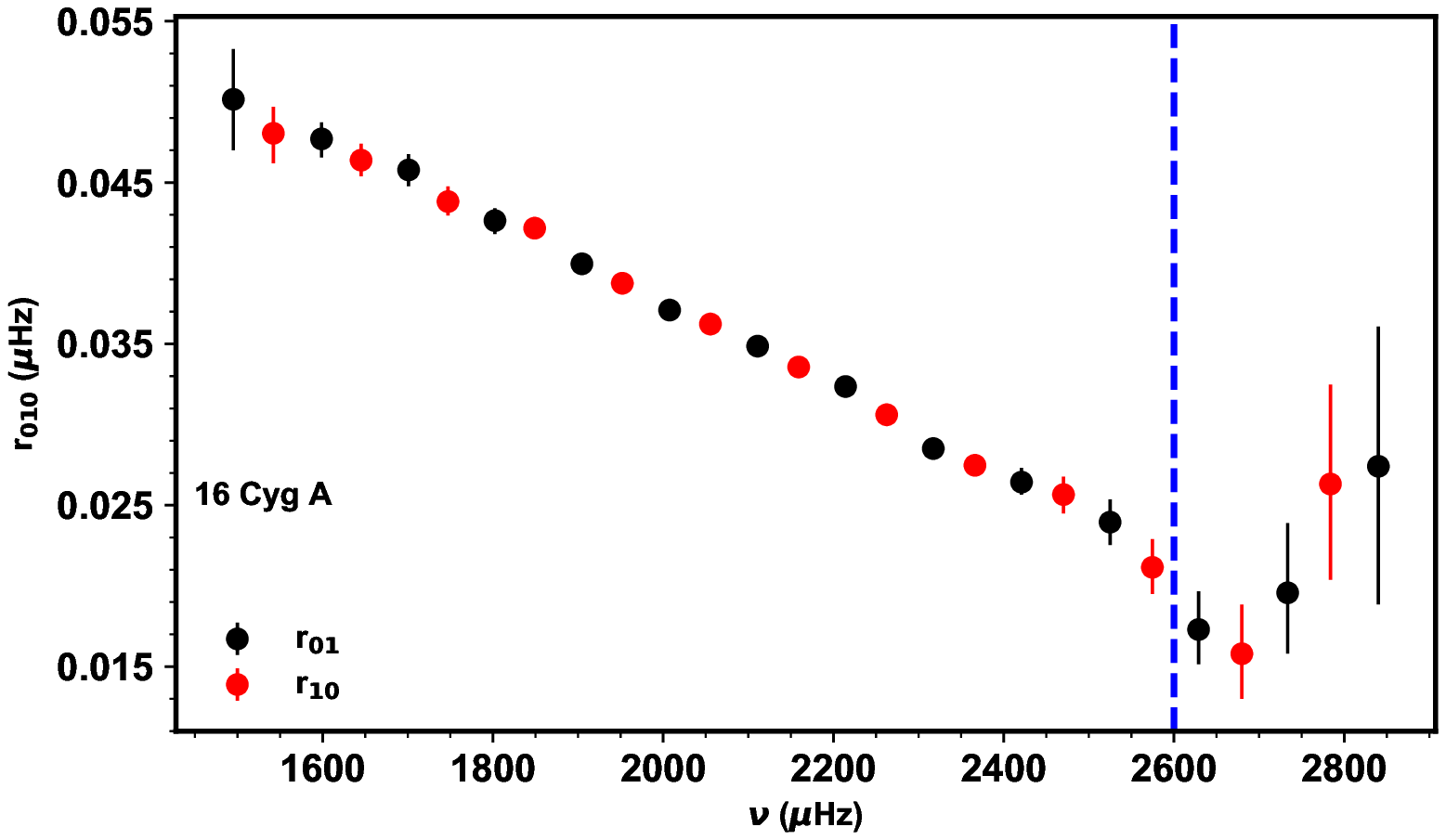}
	\quad
		\includegraphics[width=\columnwidth]{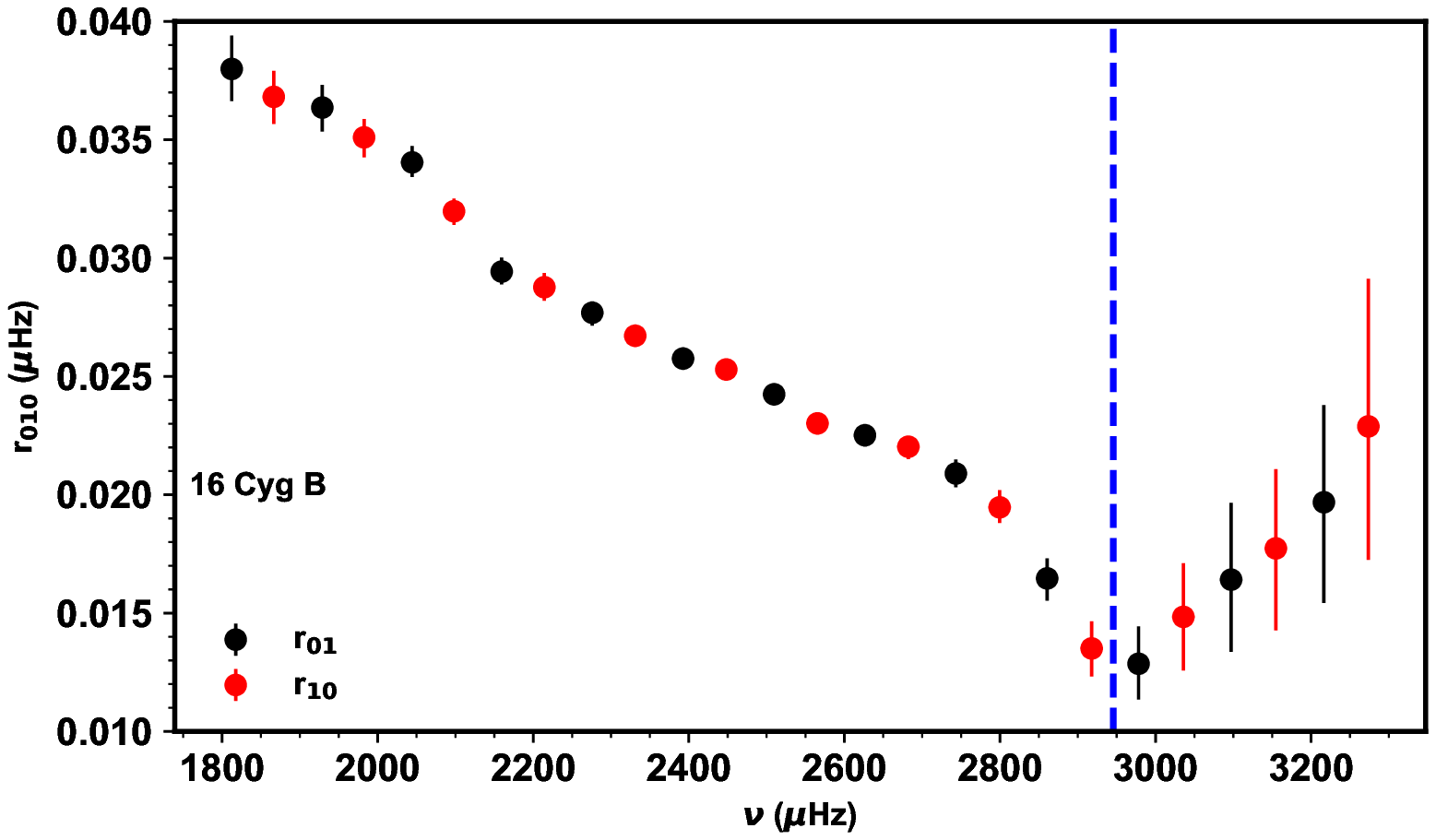}
	\quad
		\includegraphics[width=\columnwidth]{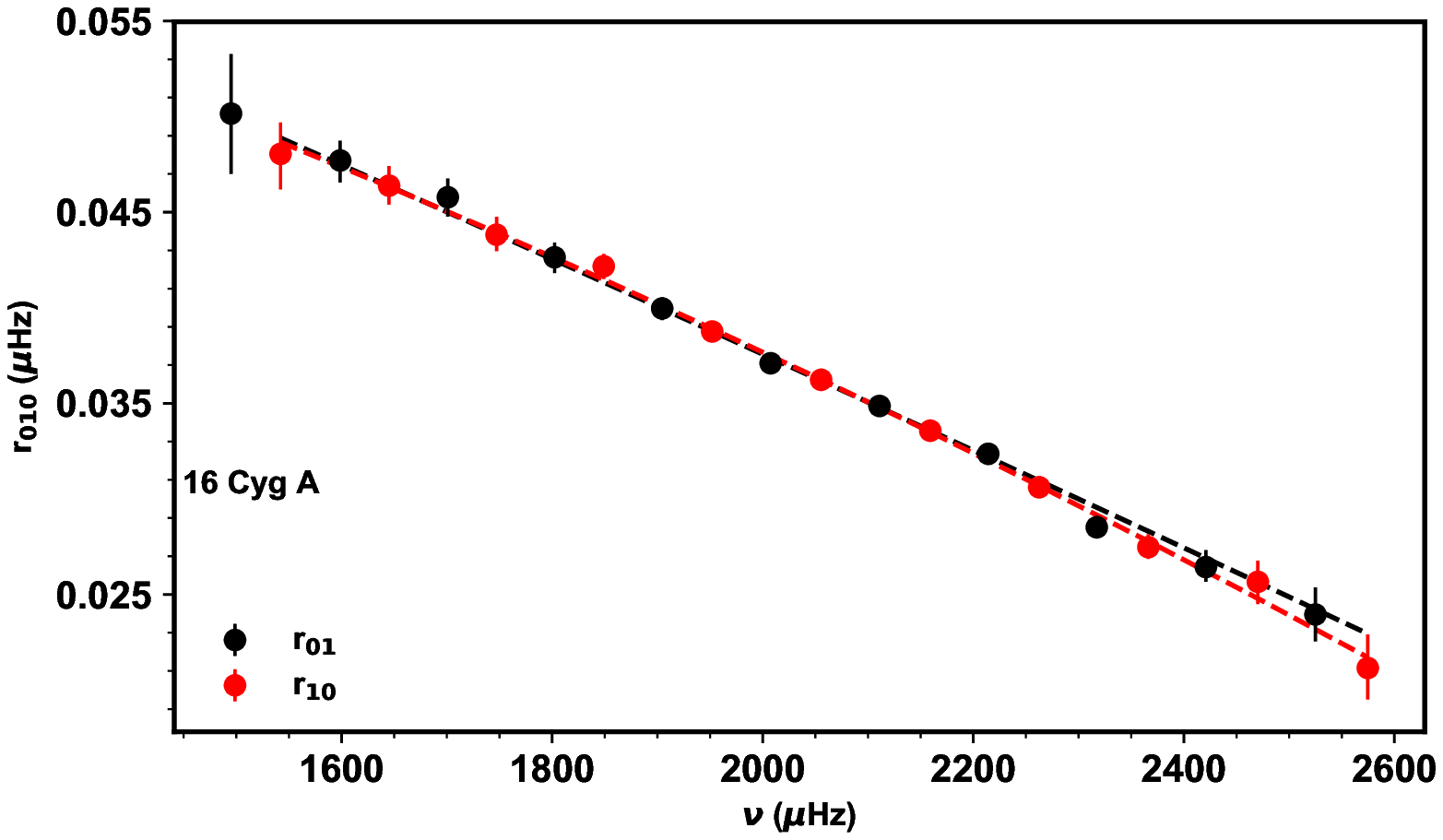}
	\quad
		\includegraphics[width=\columnwidth]{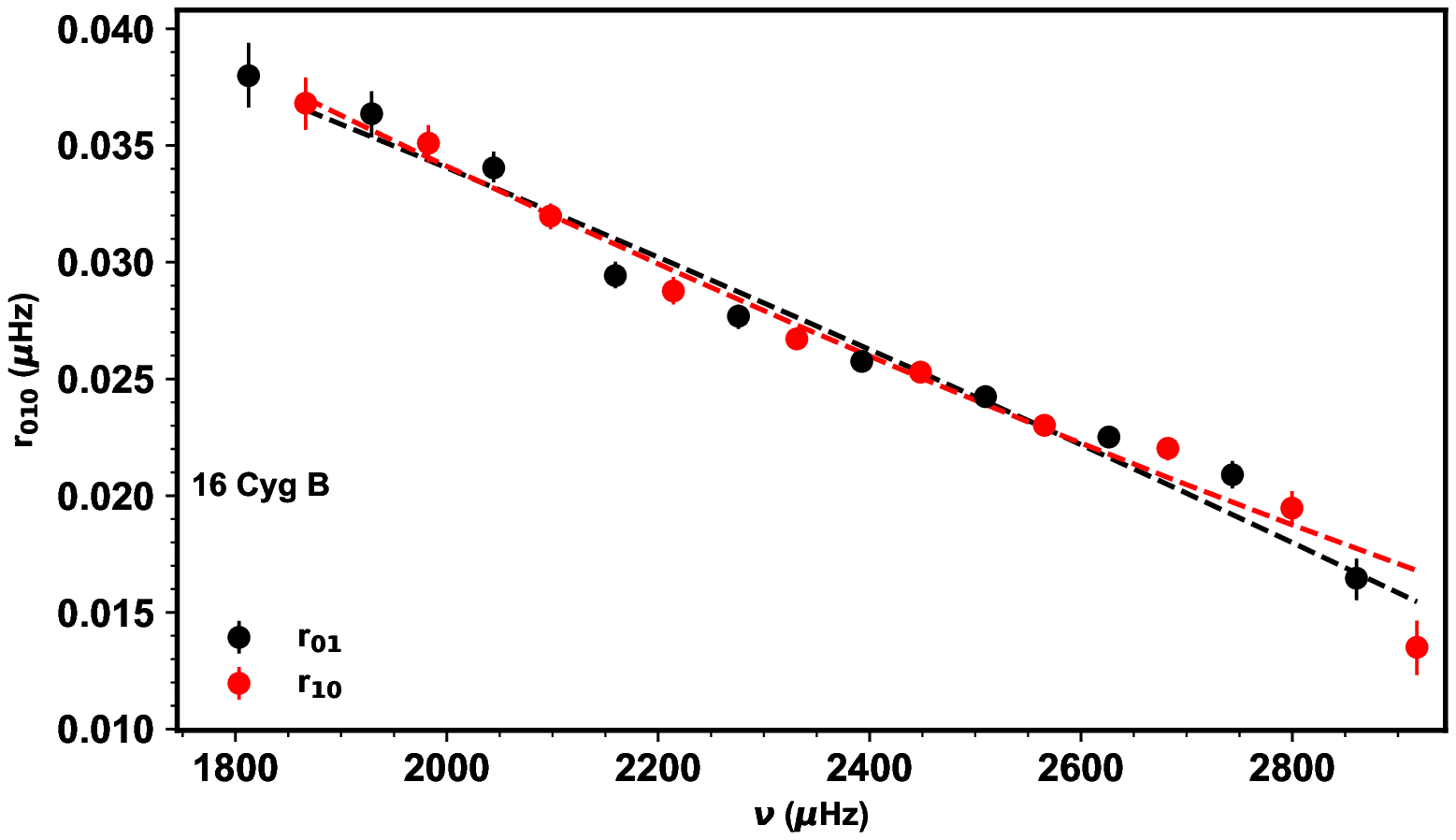}
      \caption{
      Ratios $r_{01}$ and $r_{10}$ for 16 Cyg A (top left and bottom left panels) and B (top right and bottom right panels). The blue vertical line in the top panels corresponds to the cut-off region of the ratios considered in the polynomial fits. The dashed lines in the bottom panels show the second-order polynomial fits to $r_{01}$ (black) and $r_{10}$ (red). The high-end frequencies are truncated in the bottom panels (see text for details).
      }
   \label{fits}
\end{figure*}
Here  $\beta$, $\gamma_1$, and $\gamma_2$ are chosen to ensure that  $P(\nu)$ is a sum of orthogonal polynomials, hence, that the coefficients $a_0$, $a_1$ and $a_2$, which are the parameters inferred from the fit, are uncorrelated. More recently, \cite{2020Viani} have performed similar fits to a sample of {\it{Kepler}} stars confirming a potential correlation between $a_1$ and the amount of core overshoot. 

In this study we follow \cite{2016Deheuvels} and fit a second order polynomial to the observed ratios of 16~Cyg A and B, as well as to the ratios derived for an ensemble of models representative of these stars identified through the grid-based modelling approach described in Section~\ref{models}. Nevertheless, we note below three differences between our approach and that applied in \cite{2016Deheuvels} that should be kept in mind when comparing the two studies. 

The first difference is that we fit the sets of ratios $r_{01}$ and $r_{10}$ separately. As argued by \cite{2017roxburgh,2018roxburgh}, from $N$ pairs of frequencies of $(l=0,l=1)$  modes one can only derive $N$ values of the surface-independent quantities. The $r_{01}$ and $r_{10}$ are thus highly correlated and combining them does not add any significant information. In fact, attempts to fit them simultaneously are faced with having to invert nearly singular covariance matrices. In \cite{2016Deheuvels} this problem was detected and mitigated  by applying a truncated singular value decomposition to the covariance matrix. Here, we opt, instead, to fit the two sets of ratios separately and compare the results of the two fits. The two sets of ratios computed from the individual mode frequencies of 16~Cyg A and B are shown by different colours in the top panels of Figure~\ref{fits}. Quite noticeable on the higher frequency end (to the right of the dashed blue line) is a sudden increase in the ratios. This increase is not currently reproducible by the ratios $r_{01}$ and $r_{10}$ of theoretical models. Whether this is a result of systematic errors on these higher frequencies, which have higher uncertainties, or the result of missing physics in the stellar models, it is currently unknown. Nevertheless, given that the low degree polynomial model proposed by \cite{2016Deheuvels} and used here to fit the ratios cannot capture this feature at the high-end frequency,  we truncate the observed ratios of each star at the frequency indicated by the blue dashed line when performing the fits (i.e. at $\nu_{\rm At}=$ 2600 $\mu$Hz and $\nu_{\rm Bt}=$ 2945 $\mu$Hz, respectively). Another feature that is noticeable in Figure~\ref{fits} is the signature of glitches. The signature is superimposed on the smooth behaviour of the ratios and occurs on short frequency scales. The polynomial fits to the $r_{01}$ and $r_{10}$ sets are affected slightly differently by these short-scale variations, leading to non-negligible differences in the parameters inferred from the fits. We therefore take the differences in the parameters inferred from the two fits as a measure of the uncertainty on these parameters resulting from the simplicity of the polynomial model. \


A second difference of our approach to that of \cite{2016Deheuvels} concerns the definition of the polynomial used in the fit.  With the aim of working with dimensionless parameters only, we have slightly changed the polynomial function, such that
\begin{equation}
    P(\nu) = a_0 + a_1\left(\frac{\nu}{\nu_{\rm{max}}} -\beta  \right) + a_2\left(\frac{\nu}{\nu_{\rm{max}}} - \gamma_1 \right)\left(\frac{\nu}{\nu_{\rm{max}}} - \gamma_2   \right)~,
    \label{poly_fits}
\end{equation}
where the frequency of maximum power is fixed to the observed value (i.e. 2188 $\mu$Hz and 2561 $\mu$Hz for 16 CygA and B, respectively; \citealt{2017Lund}), regardless of the fitting being performed to the observed or model ratios. Therefore, our dimensionless $a_1$ and $a_2$ values are not directly comparable with the values in \cite{2016Deheuvels} (which are expressed in mHz$^{-1}$).

The third and final difference is that we do not consider in the fit the correlation between the values of the observed ratios. These correlations are expected from the definition of the ratios (cf. Eqs~(\ref{eq1})-(\ref{eq4})) given that ratios of different $n$ share common frequencies. Instead, we use emcee algorithm in python \citep{2013Foreman}
to find the  parameters that provide the best fit of Eq.~(\ref{poly_fits}) to the observed ratios while ignoring the correlations. We then perform a set of 
10,000 
Monte Carlo simulations by perturbing the observed frequencies within their errors assuming the latter are normally distributed. For each simulation we compute new sets of ratios $r_{01}$ and $r_{10}$ and perform new fits. The median and standard deviation of the distribution for each parameter $a_0$, $a_1$ and $a_2$ are then taken as the point estimates and uncertainties. 
We note that the highest posterior density (HPD) procedures were used to determine the credible intervals, taking 68.27\% and 99.73\% for the 1$\sigma$ and 3$\sigma$ uncertainties, respectively. Briefly, HPD region of confidence interval, $\alpha$, is a ($1-\alpha$)-confidence region which satisfies the condition that the posterior density for every point in this interval is higher than the posterior density for any point outside of this interval (see \citealt{99Chen,2015Hari}). We stress that a 100($1-\alpha$)\% HPD interval is more desirable to be applied in situations where a marginal distribution is not symmetric.

Due to our neglect of the correlations between the ratios in each fit, the standard deviations derived in this manner will be an upper limit to the true formal uncertainties and can be considered as conservative errors on the parameters. The fits to the ratios are  illustrated in the lower panels of Figure~\ref{fits}. As we will see in Section~\ref{rat_X}, the error inferred on the parameters through our Monte Carlo analysis is still smaller than the difference in some of the parameters inferred from fitting the two sets of ratios ($r_{01}$ and $r_{10}$; black and red lines in  Figure~\ref{fits}, respectively), justifying further our option to simplify the procedure by not considering the correlations. 

 Prior to fitting Eq.~(\ref{poly_fits}) to the ratios, we compute $\beta$, $\gamma_1$ and $\gamma_2$ from the observations, following the procedure explained in the Appendix B of \cite{2016Deheuvels}. As we neglect the correlations between the observed ratios, $\beta$ is simply the mean of the dimensionless frequencies $\nu/\nu_{\rm max}$ considered in the fit. The observed values of $\beta$, $\gamma_1$ and $\gamma_2$, as well as the observed $\nu_{\rm max}$, are used also in the fits to the model ratios, so that the parameters inferred from fitting the model ratios can be directly compared with those derived from fitting the observations. As in \cite{2016Deheuvels}, we find that the coefficient $a_2$ is very small, so that the ratios vary nearly linearly with frequency. Hence, $a_0$ and $a_1$ can be interpreted approximately as the value of the ratio at the middle of the frequency interval considered in the fit and the slope of the linear trend, respectively.   We note that for each star, the fits to the model and both sets of observed ratios ($r_{01}$ and $r_{10}$) are all performed in the same frequency range, defined by the observations.  This is a critical point, as model and observed ratios need to be compared at the same frequency. When performing model-observation comparison of individual values of the ratios, this is assured by first interpolating the model ratios to the observed frequencies. Here, we do not compare individual ratios, but do nevertheless perform a comparison of the parameters inferred from the fit of Eq.~(\ref{poly_fits}) to the model and observed ratios. Therefore, the range of frequencies over which the fits are performed needs to be the same and it impacts directly the value of $a_0$.

\section{Results and Discussions}
\label{results}
\begin{figure}
	\includegraphics[width=\columnwidth]{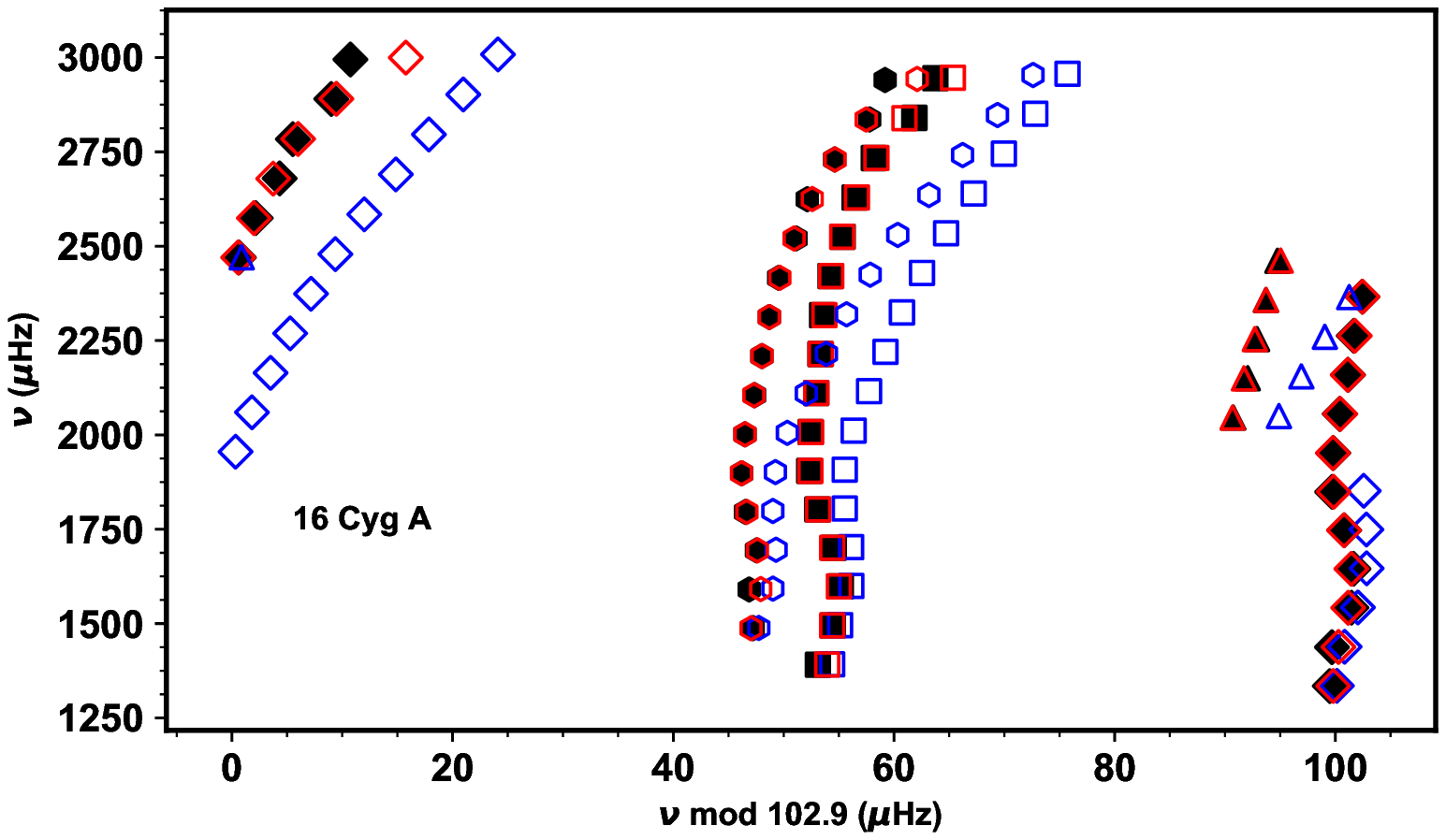}
	\quad
		\includegraphics[width=\columnwidth]{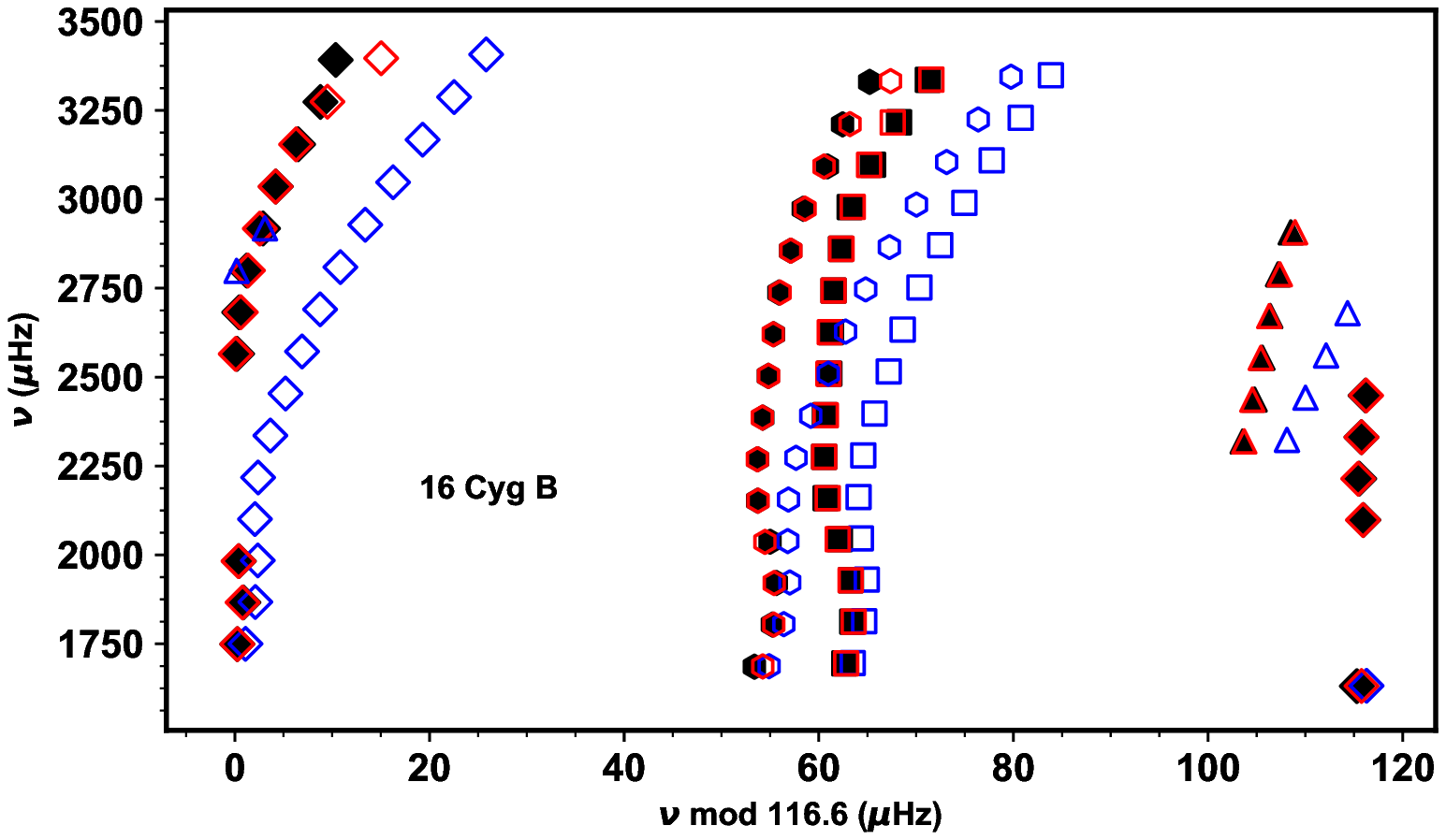}
     \caption{\'{E}chelle diagram of 16 Cyg A (top panel) and  B (bottom panel). 
     The $l = 0, 1, 2,$ and $3$ mode frequencies are represented by squares, diamonds, 	hexagons, and triangle symbols, respectively.
     The black, blue, and red symbols correspond to the observed frequencies, theoretical frequencies, and corrected frequencies of the different modes.
     }
    \label{echelle}
\end{figure}

Figure~\ref{echelle} shows a comparison of the observed frequencies with the best-fitting models of 16 Cyg A and B obtained from grid G$_{\rm free}$. The trend of the observed frequencies is well reproduced by the best-fitting model frequencies with the two-term surface correction recipe (see Eq.~\ref{correction}) taken into account. A similar trend as shown in Figure~\ref{echelle} is also observed for the best-fitting models of 16 Cyg A and B generated from grid G$_{1.4}$ and G$_{2.0}$.
We note that a systematic offset between the observed and corrected frequencies still exists for a handful of points at the highest frequency end. 
This is because high frequencies are more sensitive to the stellar outer region which is more prone to surface effects. In fact the frequency offset of up to 15 $\mu$Hz is reported at the high frequency end and the empirical corrections may not perfectly account for the differences in these regions. A detailed analysis of the robustness of the surface correction methods on the main-sequence phase has been addressed in \citet{2018Nsamba,2018Basu,2018Compton,2021Cunha} and for the evolved evolution phase in \citet{2017Ball,2020rgensen,2021Ong}.

\subsection{Inferred stellar parameters of 16 Cyg A and B}
\label{parameters}
Grid G$_{1.4}$ and G$_{2.0}$ employ a fixed helium-to-heavy element enrichment ratio (see Table~\ref{grids}) via Eq.~(\ref{law}), used to determine how the initial helium mass fraction changes with metal mass fraction. \citet{2019Verma} demonstrated using a subset of {\it{Kepler}} ``Legacy'' sample stars that the scatter in the relation between initial helium mass fraction and metal mass fraction is significant, rendering this relation unsuitable for single star studies, especially in the case of population I stars. With this in mind, we consider grid G$_{\rm free}$ as the chosen reference grid since no initial helium restriction is set in this grid other than a lower and upper limit (see Section~\ref{models}).
Recently, \citet{2021Nsamba} reported that the inferred stellar masses and radii from grids with a fixed helium-to-heavy element enrichment ratio are systematically lower than those from grids with free initial helium mass fraction. Similar findings were reached by \citet{2021Deal}, while consistent values were found by the same authors for the central hydrogen mass fractions and ages. Therefore, grid G$_{1.4}$ and G$_{2.0}$ are employed here to verify the consistency of our findings concerning these parameters when frequency ratios are applied to tightly constrain the central hydrogen mass fractions and ages of 16 Cyg A and B (see Section~\ref{rat_X}).

Table~\ref{grids2} and Table~\ref{grids3} show the parameters inferred in our study, and their corresponding 1$\sigma$ uncertainties, for 16 Cyg A and B, respectively. 
The stellar parameters missing in these tables are not provided because they are not available in the model files of these grids.
\begin{table*}
\centering
\caption{Derived stellar parameters and their associated 1$\sigma$ uncertainties for 16 Cyg A.}
\label{grids2}
\begin{tabular}{c c ccccccc }        
\hline 
Grid name   & $M (\rm{M}_\odot$) & $R (\rm{R}_\odot$) & Age (Gyr) & $X_{\rm c}$ & $\alpha_{\rm{mlt}}$ & $Z_{\rm 0}$ & $L (\rm{L_\odot})$ & $Y_{\rm{surf}}$\\
\hline \hline
G$_{1.4}$   & 1.08 $\pm$ 0.02 & 1.226 $\pm$ 0.010 & 6.5 $\pm$ 0.3 & 0.04 $\pm$ 0.02  &  1.94 $\pm$ 0.06   & 0.026 $\pm$ 0.002 & 1.62 $\pm$ 0.06  & -    \\
G$_{2.0}$   & 1.05 $\pm$ 0.02   & 1.213 $\pm$ 0.008 & 6.4 $\pm$ 0.3 & - & 1.86 $\pm$ 0.06 & 0.026 $\pm$ 0.001 & 1.55 $\pm$ 0.05 & -        \\
G$_{\rm{free}}$   &   1.09 $\pm$ 0.03 &   1.233 $\pm$ 0.010 &     6.5 $\pm$ 0.2  &    0.02 $\pm$ 0.01  & 1.90 $\pm$ 0.07   & 0.026 $\pm$ 0.001 & 1.60 $\pm$ 0.06   &  0.229 $\pm$ 0.011   \\

\hline                                 
\end{tabular}
\end{table*}
\begin{table*}
\centering
\caption{Derived stellar parameters and their associated 1$\sigma$ uncertainties for 16 Cyg B.}
\label{grids3}
\begin{tabular}{c c ccccccc }        
\hline 
Grid name   & $M (\rm{M}_\odot$) & $R (\rm{R}_\odot$) & Age (Gyr) & $X_{\rm c}$  & $\alpha_{\rm{mlt}}$ & $Z_{\rm 0}$ & $L (\rm{L_\odot})$ & $Y_{\rm{surf}}$\\
\hline \hline
G$_{1.4}$   & 1.01 $\pm$ 0.02 & 1.102 $\pm$ 0.010 & 7.4 $\pm$ 0.3 & 0.16 $\pm$ 0.02 & 1.81 $\pm$ 0.06  & 0.023 $\pm$ 0.002 & 1.18 $\pm$ 0.04 & -        \\
G$_{2.0}$   & 0.98 $\pm$ 0.02 & 1.094 $\pm$ 0.008 & 7.3 $\pm$ 0.3 & - & 1.79 $\pm$ 0.04 & 0.022 $\pm$ 0.001 & 1.18 $\pm$ 0.04  & -        \\
G$_{\rm{free}}$   &  1.03 $\pm$ 0.03 &    1.111 $\pm$ 0.010  & 7.3 $\pm$ 0.3 &    0.14 $\pm$ 0.02   & 1.84 $\pm$ 0.07     & 0.022 $\pm$ 0.001  & 1.19 $\pm$ 0.05  &   0.225 $\pm$ 0.012         \\

\hline                                 
\end{tabular}
\end{table*}
\begin{figure*}
	\includegraphics[width=\columnwidth]{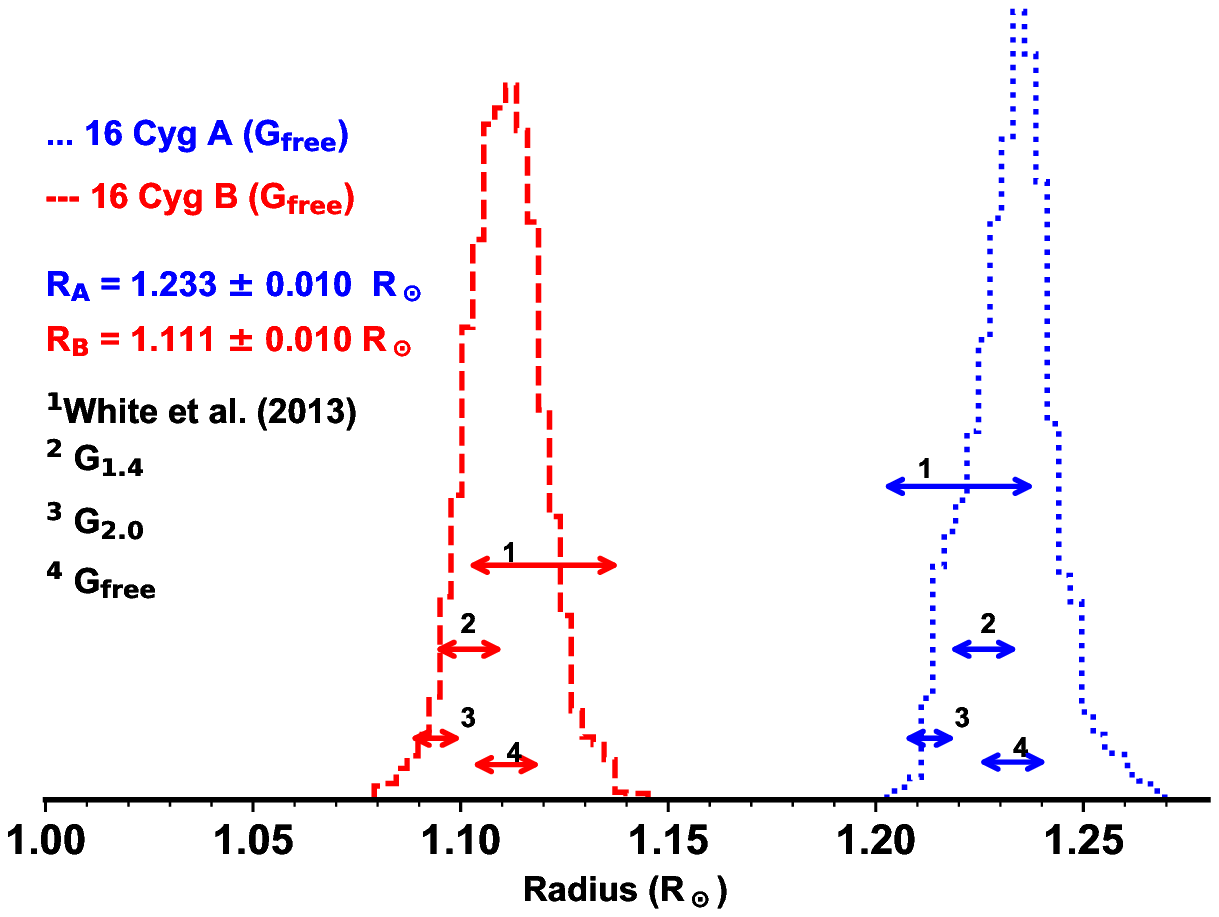}
	\quad
		\includegraphics[width=\columnwidth]{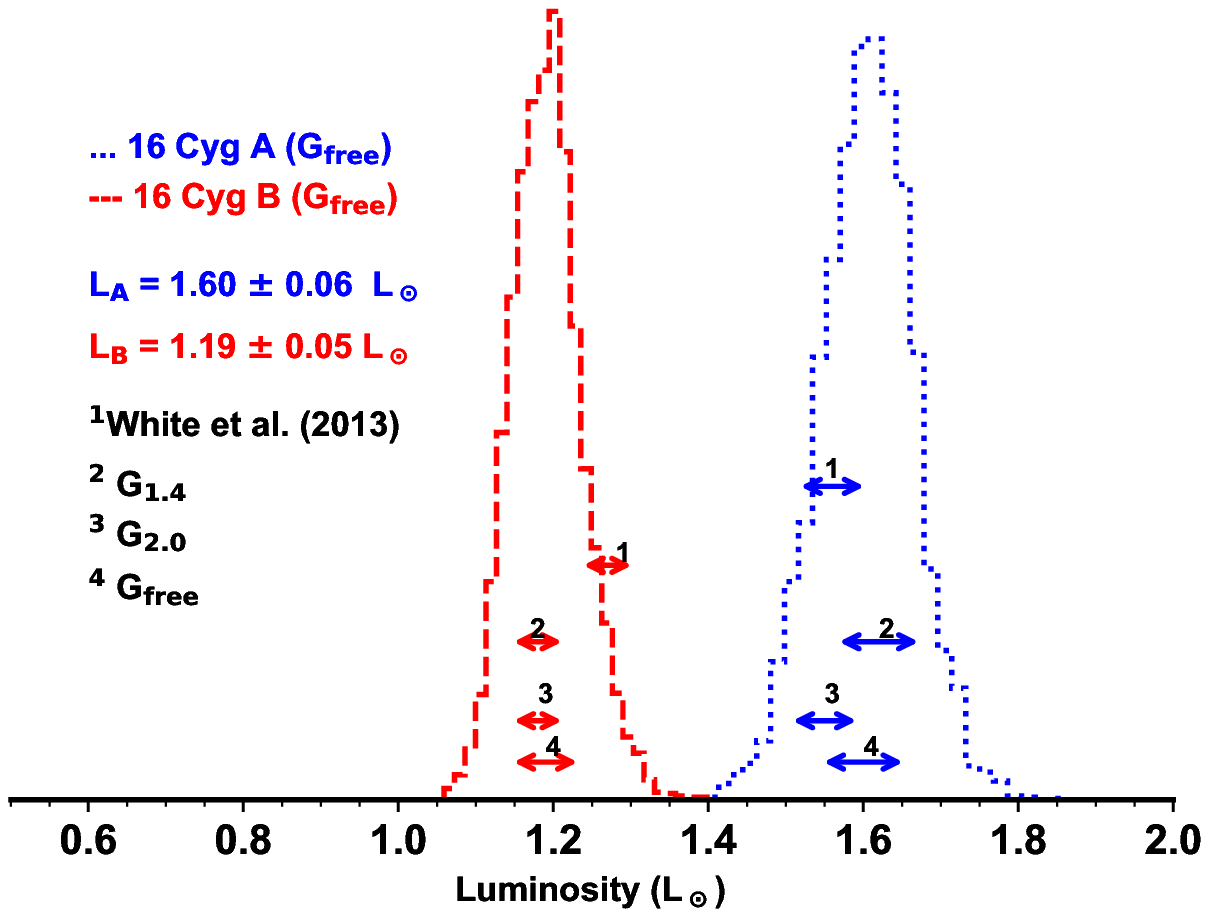}
		\includegraphics[width=\columnwidth]{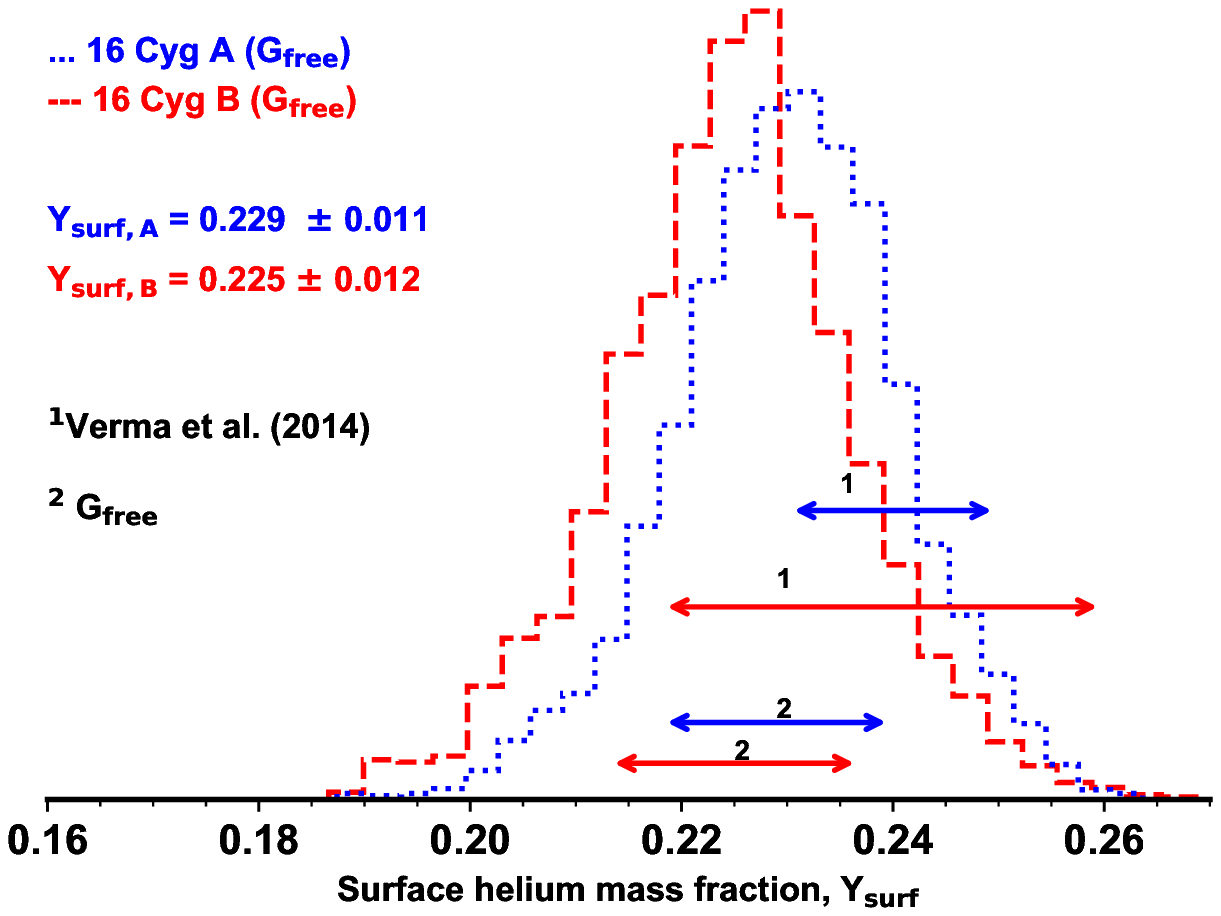}
	\quad

    \caption{Probability densities of inferred parameters of 16 Cyg A (dotted blue lines) and B (dashed red lines) from grid G$_{\rm free}$ and corresponding mean values and standard deviations. Horizontal arrows show 1$\sigma$ range of parameters inferred from interferometric measurements \citep{2013White}, glitch analysis \citep{2014Verma}, grid G$_{1.4}$, G$_{2.0}$, and G$_{\rm{free}}$ according to the numbers' description in each panel.
    }
   \label{model_independent}
\end{figure*}
\begin{figure*}
		\includegraphics[width=\columnwidth]{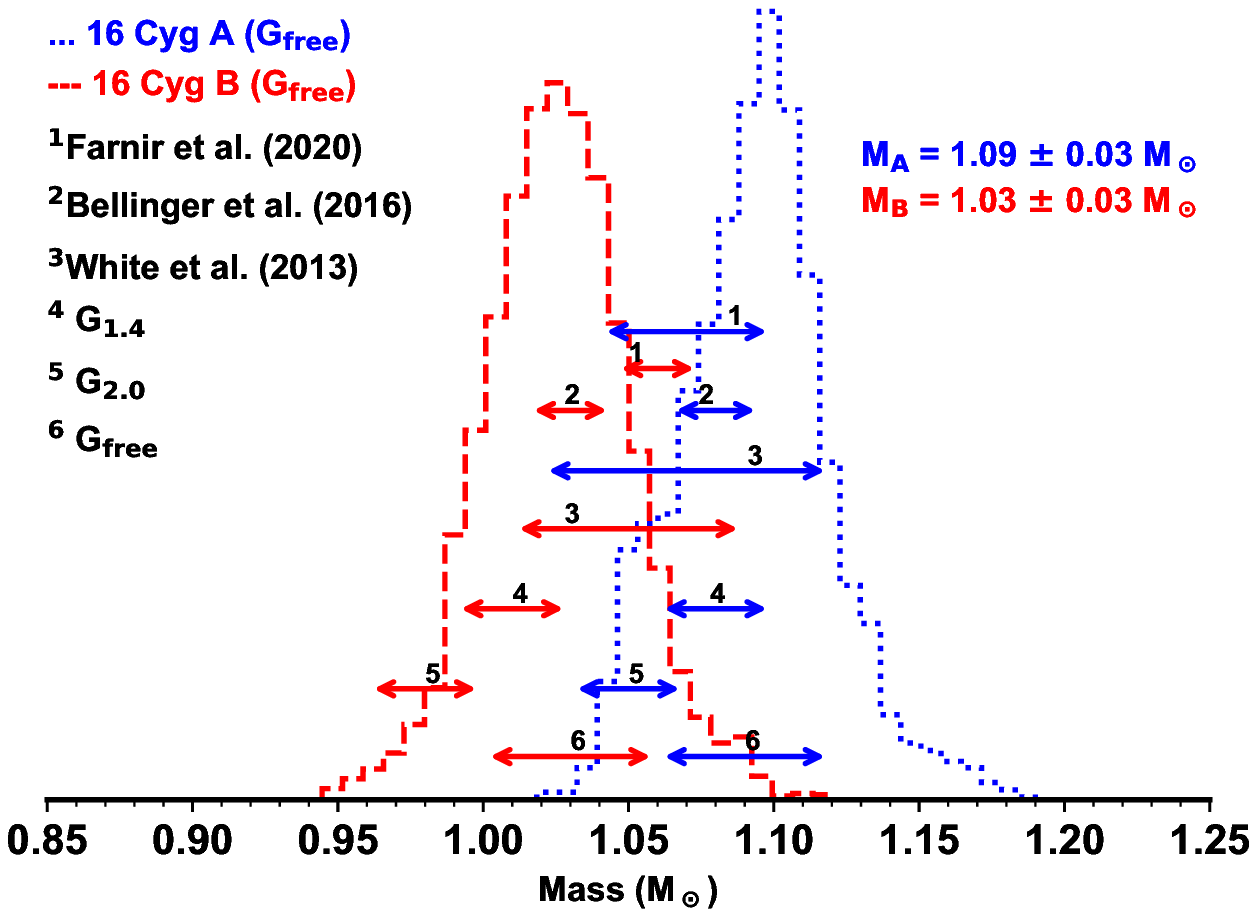}
	\quad
		\includegraphics[width=\columnwidth]{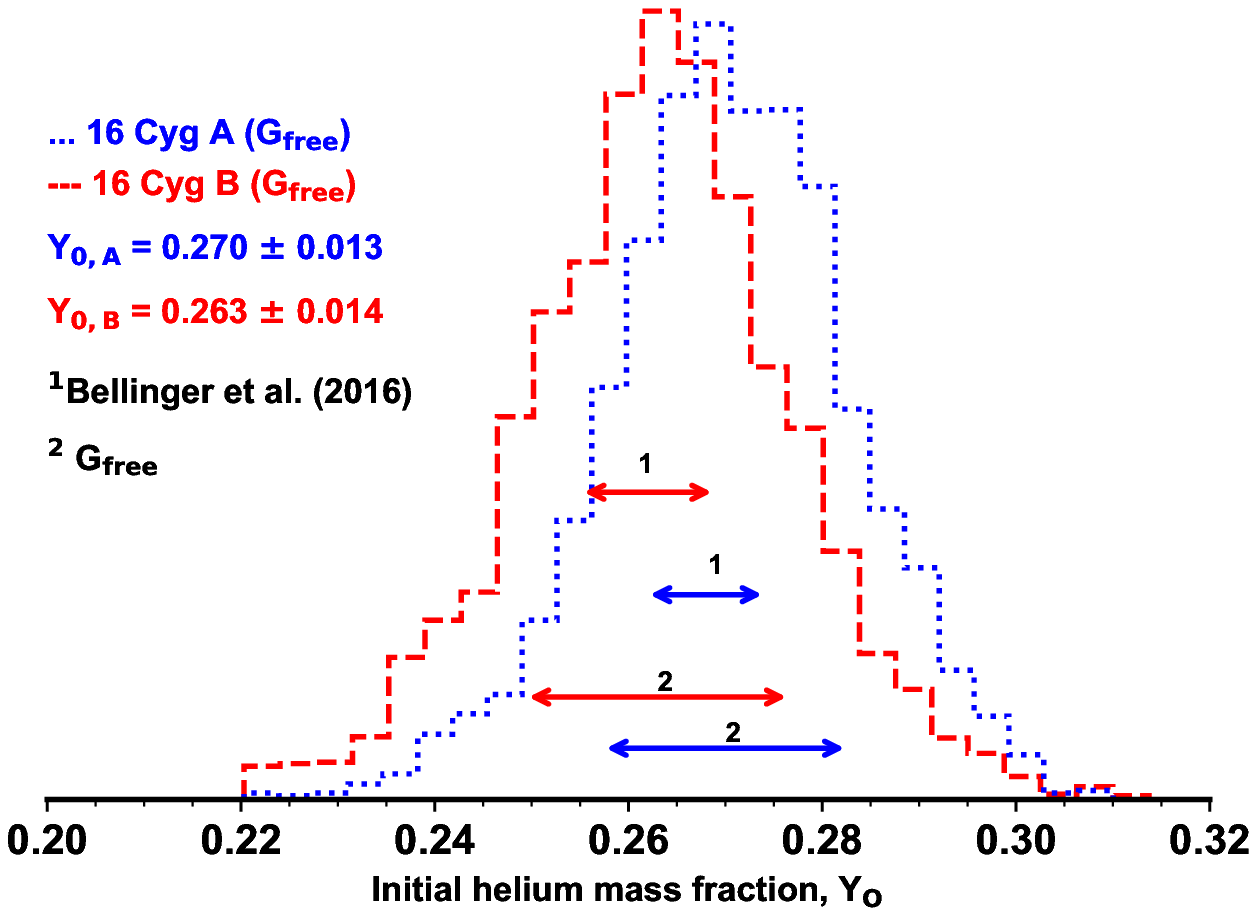}
	\quad
		\includegraphics[width=\columnwidth]{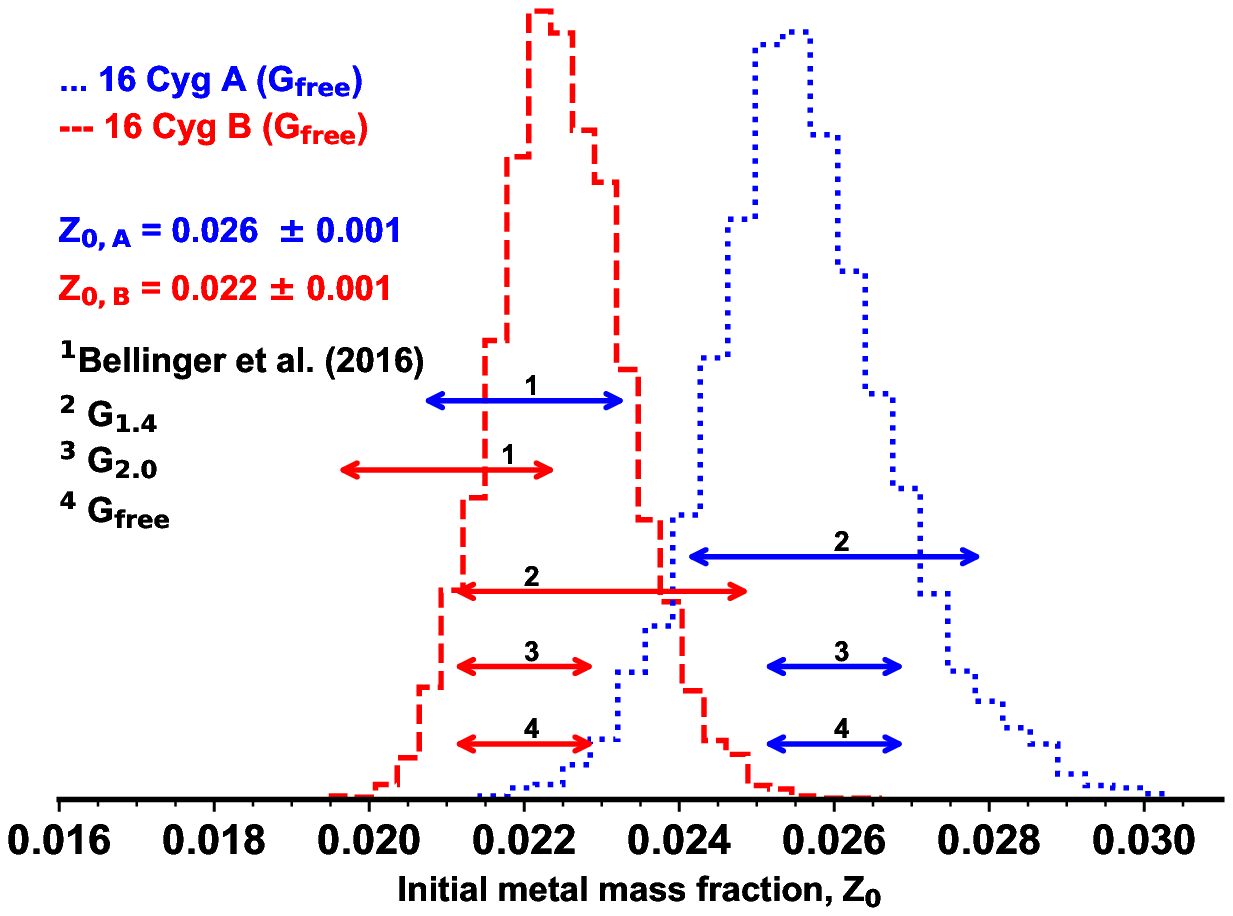}
	\quad
		\includegraphics[width=\columnwidth]{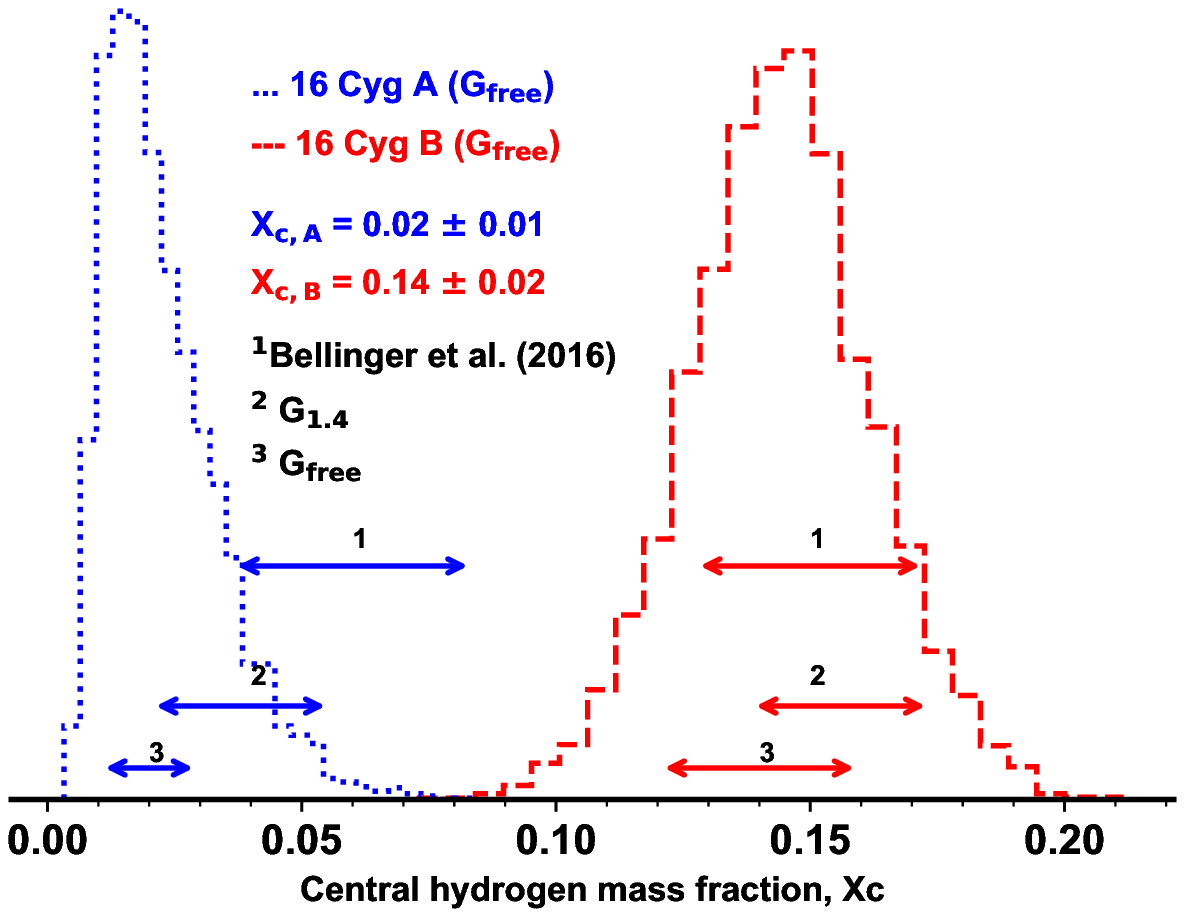}
	\quad
		\includegraphics[width=10cm,height=6.5cm]{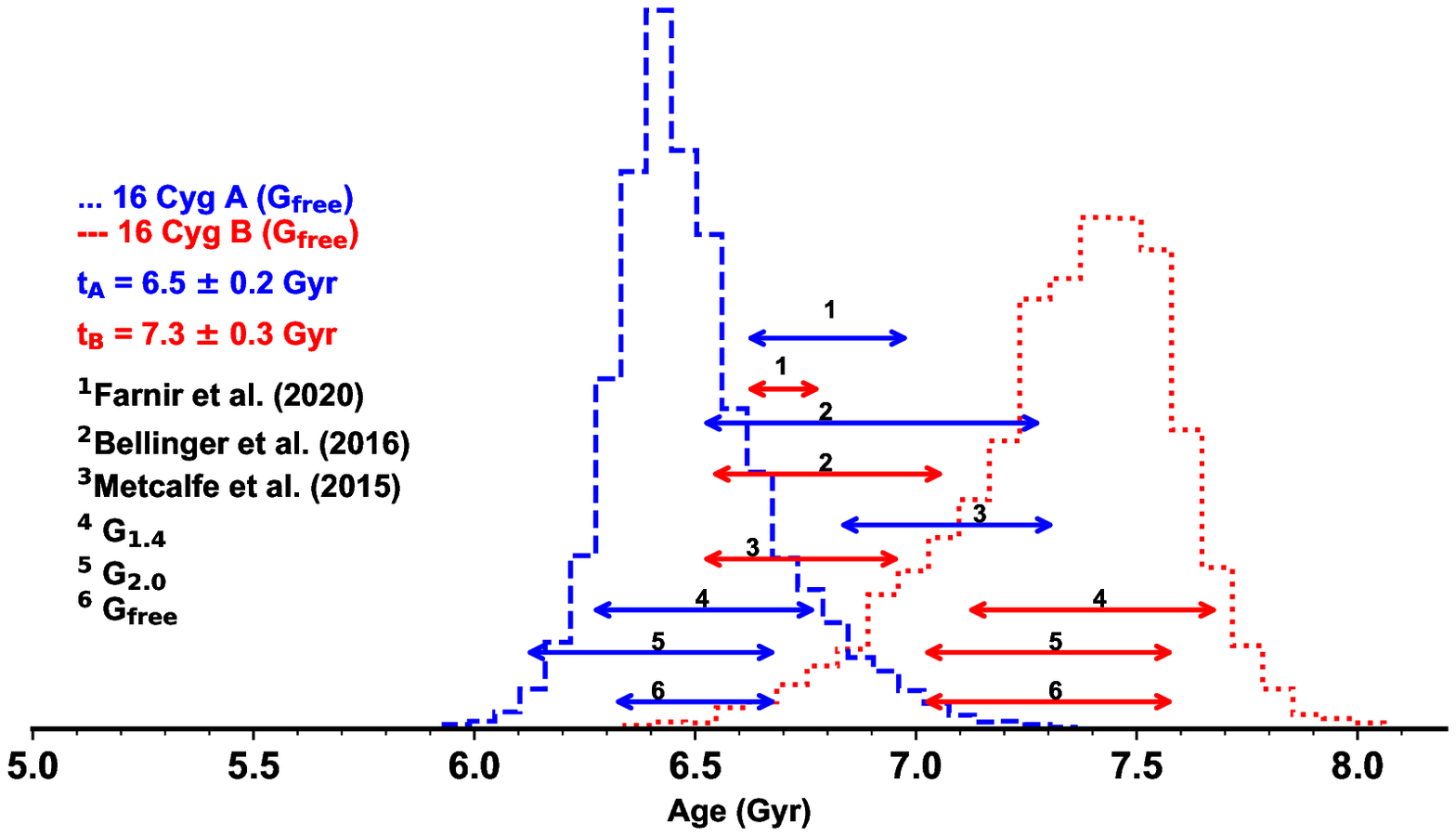}
 \caption{Probability densities of inferred parameters of 16 Cyg A (dotted blue lines) and B (dashed red lines) from grid G$_{\rm free}$ and corresponding mean values and standard deviations. Masses (top left), initial helium mass fractions (top right), initial metal mass fractions (middle left), central hydrogen mass fractions (middle right), and ages (bottom panel). Horizontal arrows show 1$\sigma$ range of predicted parameters from \citet{2013White}, \citet{2016Bellinger}, \citet{2015Metcalfe}, \citet{2020Farnir}, grid G$_{1.4}$, G$_{2.0}$, and G$_{\rm{free}}$ according to the numbers' description in each panel.
   }
   \label{model_parameters}
\end{figure*}

Prior to applying the ratios $r_{01}$ and $r_{10}$ (following a description in Section~\ref{ratios}) as a diagnostic of the interior characteristics of the best-fit models of 16 Cyg A and B, we explore the consistency of the inferred parameters from the employed model grids with literature findings. 
A detailed comparison between the stellar parameters inferred from grid G$_{1.4}$ and G$_{2.0}$ (i.e., employing a fixed value of enrichment ratio) with the reference grid G$_{\rm free}$ (i.e., free initial helium abundance), and literature findings is shown in Figure~\ref{model_independent} and Figure~\ref{model_parameters}.
The inferred radii and luminosities of 16 Cyg A and B from grids  G$_{1.4}$, G$_{2.0}$, and G$_{\rm free}$ agree within 1-2$\sigma$ uncertainties with the model independent radii and luminosities from \citet{2013White}, respectively. This is shown in the top panels of Figure~\ref{model_independent}. 
We note that \citet{2013White} derived the stellar radius, $R$, through a combination of the angular diameter, $\theta_{\rm{LD}}$, with parallax-based distance to the star, $D$, using the expression
\begin{equation}
    R = \frac{1}{2}\theta_{\rm{LD}}D~.
\end{equation}
They determined the effective temperature, $T_{\rm{eff}}$ of 16 Cyg A and B using a bolometric flux (\citealt{Boya}) at the Earth, $F_{\rm{bol}}$ via the relation
\begin{equation}
    T_{\rm{eff}} = \left(\frac{4F_{\rm{bol}}}{\sigma \theta_{\rm{LD}}^2}   \right)~.
\end{equation} 
Similarly, the stellar luminosities were calculated from the bolometric flux and parallax-based distances.
The bottom panel of Figure~\ref{model_independent} shows an excellent agreement between the inferred surface helium mass fractions from grid G$_{\rm free}$ and values reported by \citet{2014Verma} using glitch analysis.

The top left panel of Figure~\ref{model_parameters} shows a comparison of the derived masses of 16 Cyg A and B from our grids with those estimated from \citet{2013White}, \citet{2016Bellinger}, and \citet{2020Farnir}. The masses of both 16 Cyg A and B inferred from all the grids agree within 1-2$\sigma$ uncertainties with literature findings. \citet{2013White} reports the largest uncertainties on the masses of 16 Cyg A and B. This is because they estimated the stellar masses  based on scaling relations between
the large frequency separation of solar-like oscillations, $\Delta \nu$, and the density of the star \citep{1986Ulrich}, which take the form:
\begin{equation}
    \frac{\Delta \nu}{\Delta \nu_\odot} = \left( \frac{M}{M_\odot} \right)^{(1/2)} \left(\frac{R}{R_\odot} \right)^{(-3/2)} ~,
    \label{scaling}
\end{equation}
where $R$ is deduced from the angular diameter measurement. \citet{2020Farnir} employed a method (i.e. WhoSGLAd; \citealt{2019Farnir}) which generates only a handful of best-fit models, from which stellar parameters and their corresponding uncertainties are deduced. An interesting aspect of using ``WhoSGLAd'' optimisation tool involves its capability of exploring a set of seismic indicators including acoustic glitches, so as to generate models that are representative of the stellar structure. We note that \citet{2020Farnir} carried out numerous simulations varying the model input physics and observable constraints of 16 Cyg A and B. The results presented here are from their ``table 4'', which they obtained by varying various stellar parameters including $T_{\rm{eff}}$ and implementing a turbulent mixing with a coefficient of $D_{\rm{turb}}$ = 7500 cm$^{2}s^{-1}$. 
The results reported in table 4 of \citet{2020Farnir} satisfy their specified seismic data, interferometric radii, luminosities, effective temperatures, and surface helium abundances of both 16 Cyg A and B included in their optimisation process, except for the spectroscopic metallicities.
Our non-seismic constraints, i.e., $T_{\rm{eff}}$ and [Fe/H], were satisfied by our best-fitting models of both 16 Cyg A and B.
This also explains the slight differences in the stellar parameters derived from this work and those reported by \citet{2020Farnir}.
\citet{2020Farnir} also considered 1$\sigma$ uncertainties on the non-seismic parameters while we considered 3$\sigma$ uncertainties. This further explains the differences in the 1$\sigma$ uncertainty sizes of our results and those of \citet{2020Farnir}.
\citet{2016Bellinger} employed an approach based on machine learning to estimate the fundamental stellar parameters of 16 Cyg A and B from a set of classical and asteroseismic observations, i.e. effective temperatures, surface gravities, and metallicities from \citet{2009Ram}; radii and luminosities from \citet{2013White}; and frequencies from \citet{2015Davies}. This method involves developing a multiple-regression model capable of characterising observed stars. The construction of such a model requires developing a matrix of evolutionary simulations and employing a machine learning algorithm to unveil the relationship between model quantities and observable quantities of targeted stars (see \citealt{2016Bellinger} for details). It is also worth noting that \citet{2016Bellinger} included various variable model parameters in their grid, i.e. mass, chemical composition, mixing length parameter, overshoot coefficient, $\alpha_{\rm{ov}}$, and the diffusion multiplication factor, $D$, which they find to linearly decrease with mass. Our grids use a constant factor, i.e., $D = 1$.

Despite the agreement in the masses and radii of 16 Cyg A and B from all our grids being recorded within 1$\sigma$ uncertainties (see Table~\ref{grids2} and Table~\ref{grids3}), it can be seen that the central values of the masses and radii from grid G$_{1.4}$  are higher than those from grid G$_{2.0}$ . This stems from the difference in the helium-to-heavy element ratio values adopted in the two grids, i.e., $\Delta Y/\Delta Z$ = 1.4 and 2.0 for grid G$_{1.4}$ and G$_{2.0}$, respectively. A comprehensive review on the differences and systematic uncertainties on the fundamental stellar parameters arising from using an enrichment ratio, and including the initial helium mass fraction as a free variable in stellar grids is given in \citet{2021Nsamba}.

Again, there is a 1$\sigma$ agreement between the initial helium mass fractions (top right panel of Figure~\ref{model_parameters}) and central hydrogen mass fractions (middle right panel of Figure~\ref{model_parameters}) obtained from our grids and the results from \citet{2016Bellinger}. Concerning the hypothesis that 16 Cyg A and B were formed at the same time and with the same initial composition (conditions that were not assumed from the outset in this work), grid G$_{1.4}$  yields initial metal mass fractions which agree within 1$\sigma$ uncertainties, while grid G$_{\rm free}$ yields values that agree with 3$\sigma$ uncertainties. We find an excellent agreement between the initial metal mass fractions of 16 Cyg B reported by \citet{2016Bellinger} and those from our grids, while those of 16 Cyg A agree only within 2$\sigma$ uncertainties (middle left panel of Figure~\ref{model_parameters}).
The stellar ages inferred from our grids only conform to the hypothesis that 16 Cyg A and B are coeval within 2-3$\sigma$ uncertainties (bottom panel of Figure~\ref{model_parameters}).  It is worth noting that the results from \citet{2020Farnir} and \citet{2019Bellinger} reported in this work are those derived without imposing a common age and initial composition restrictions.
We highlight that \citet{2015Metcalfe} used the Asteroseismic Modeling Portal (AMP; \citealt{Met2009}) on a grid of models which did not include overshoot, but considered diffusion and settling of helium using the prescription of   
\citet{Michaud}. The selection of best-fit models was carried out via a $\chi^2$ based on a set of seismic and spectroscopic constraints, but considering 3$\sigma$ uncertainties on the spectroscopic parameters.
We note that during the investigation of the reliabilty of asteroseismic inferences, \citet{2015Metcalfe} carried out different combinations of observational constraints. The results of \citet{2015Metcalfe} reported in this article are those derived when taking into account all the observable constraints of 16 Cyg A and B, i.e. seismic and spectroscopic constraints.
In addition, \citet{2020Farnir} reported stellar ages of 16 Cyg A and B derived by varying a series of model input physics, such as Solar metallicity mixtures, opacities, overshoot, undershoot, turbulent mixing coefficients, element diffusion, and surface correction options, as well as imposing different treatments like different choices of seismic and atmospheric constraints in the optimisation process and restrictions (e.g. with and without common age and initial composition) .
They report the ages of 16 Cyg A and B to span the range [6.2 - 7.8] Gyr. These results are in agreement with the values returned from our grids. We now down select the best-fit models from our grids  to the ones that predict ratios $r_{01}$ and $r_{10}$ comparable to the observed values.

\subsection{Ratios $r_{01}$ and $r_{10}$ as a diagnostic for central hydrogen abundances}
\label{rat_X}
To further constrain the central hydrogen abundance of 16 Cyg A and B, we down select the set of best-fit models obtained following the forward modelling routine described in Section~\ref{models} and used to estimate the stellar parameters shown in Section~\ref{parameters}.
We perform a second order polynomial fit to the ratios $r_{01}$ and $r_{10}$ derived for the best-fit models and for the observations, for both our target stars (following the description in Section~\ref{ratios}), extracting the first two polynomial coefficients $a_{0}$ and $a_{1}$. Table~\ref{rati_1} shows the $a_{1}$ and $a_{0}$ values and their corresponding 3$\sigma$ uncertainties deduced from the second order polynomial fit to the observed ratios $r_{01}$ and $r_{10}$ of 16 Cyg A. Similar results are shown in Table~\ref{rati_2} for 16 Cyg B. Comparison of the values of $a_{1}$ inferred from fitting  $r_{01}$ and $r_{10}$ show that their difference is larger than the 1$\sigma$ uncertainties on this coefficient, both in the case of 16 Cyg A and B. For $a_{0}$, the difference between the two values is comparable to the 1$\sigma$ uncertainties in the case of 16 Cyg A, but is much larger than the 1$\sigma$ uncertainties in the case of 16 Cyg B. In practice, to account for the added uncertainty brought by the differences in the two measurements of $a_{0}$ and $a_{1}$, in our down selection we will consider all best-fit models that are within 3$\sigma$ uncertainties of either values derived for these parameters.  
\begin{table}
\centering
\caption{($a_{1}, a_{0}$) values and their associated lower and upper 3$\sigma$ uncertainties determined from the observed ratios $r_{01}$ and $r_{10}$ of 16 Cyg A.  }
\label{rati_1}
\begin{tabular}{c c c c c c c}        
\hline 
Ratios    & $a_{1}$ & $\epsilon_{-a_{1}}$   &      $\epsilon_{+a_{1}}$  &  $a_{0}$ &  $\epsilon_{-a_{0}}$  & $\epsilon_{+a_{0}}$         \\
\hline \hline
$r_{01}$    &   -0.0563 &   0.0028 & 0.0029 &   0.0372  & 0.0003 &  0.0004  \\

$r_{10}$    &   -0.0548 &   0.0025  &   0.0023  &  0.0373   &   0.0002  &   0.0003 \\ 

\hline                                 
\end{tabular}
\end{table}

\begin{table}
\centering
\caption{($a_{1}, a_{0}$) values and their associated lower and upper 3$\sigma$ uncertainties determined from the observed ratios $r_{01}$ and $r_{10}$ of 16 Cyg B.  }
\label{rati_2}
\begin{tabular}{c c c c c c c}        
\hline 
Ratios    & $a_{1}$ & $\epsilon_{-a_{1}}$   &      $\epsilon_{+a_{1}}$  &  $a_{0}$ &  $\epsilon_{-a_{0}}$  & $\epsilon_{+a_{0}}$         \\
\hline \hline
$r_{01}$    &   -0.0501 &   0.0021 & 0.0024 &   0.0275  & 0.0002 &  0.0002  \\

$r_{10}$    &   -0.0513 &   0.0020  &   0.0024  &  0.0263   &   0.0003  &   0.0002 \\ 

\hline                                 
\end{tabular}
\end{table}

\begin{figure*}
	\includegraphics[width=\columnwidth]{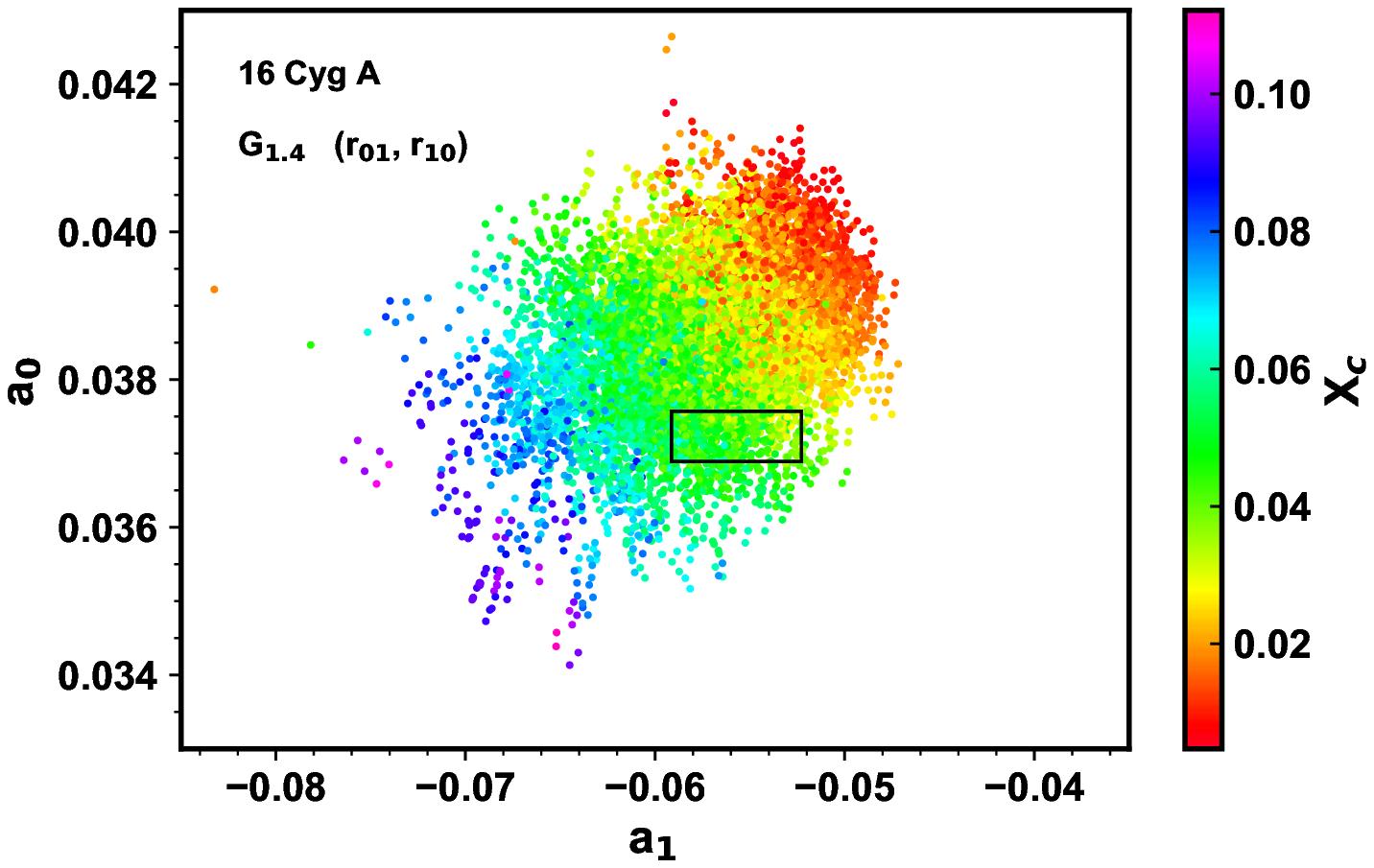}
	\quad
		\includegraphics[width=\columnwidth]{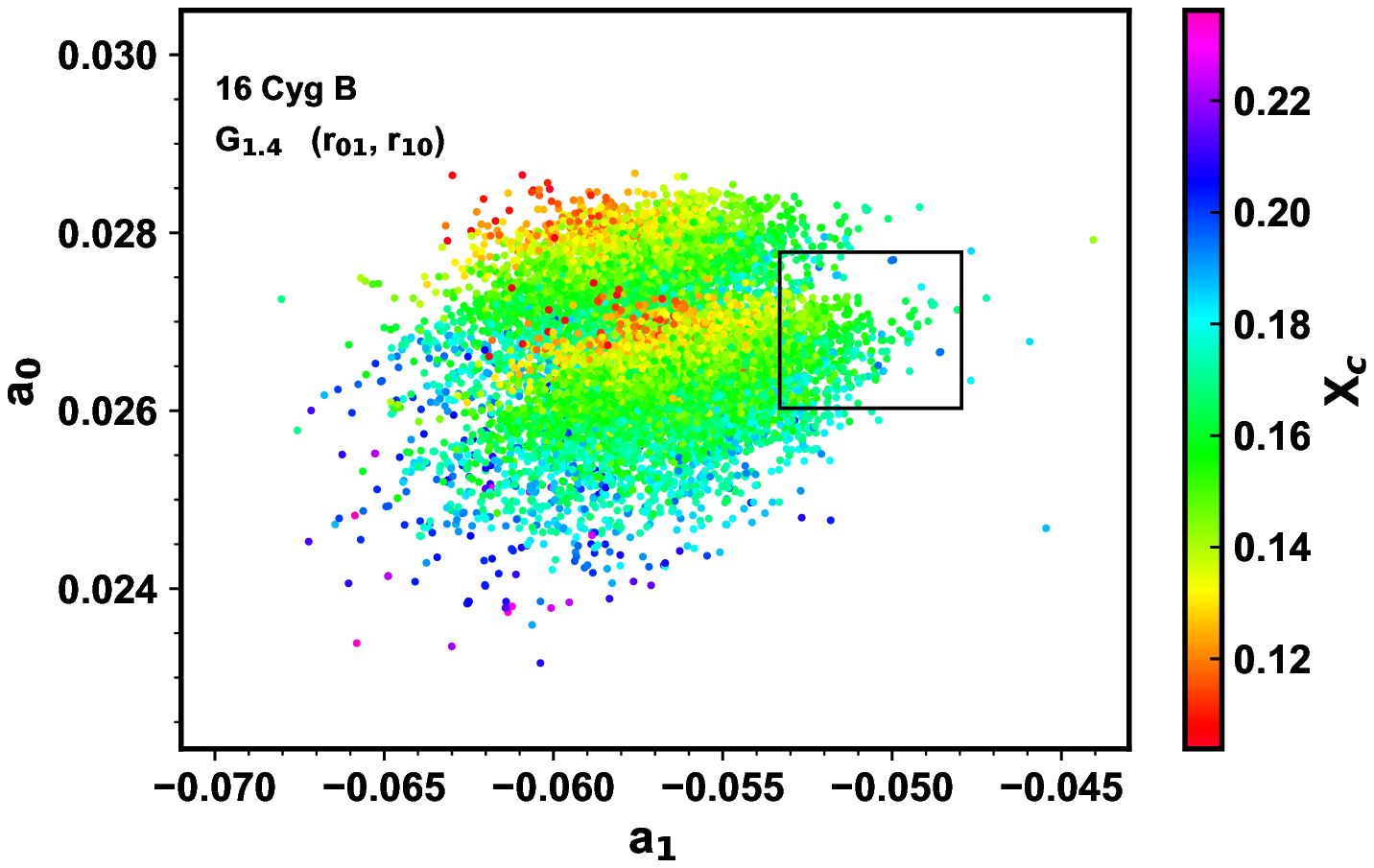}
	\quad
		\includegraphics[width=\columnwidth]{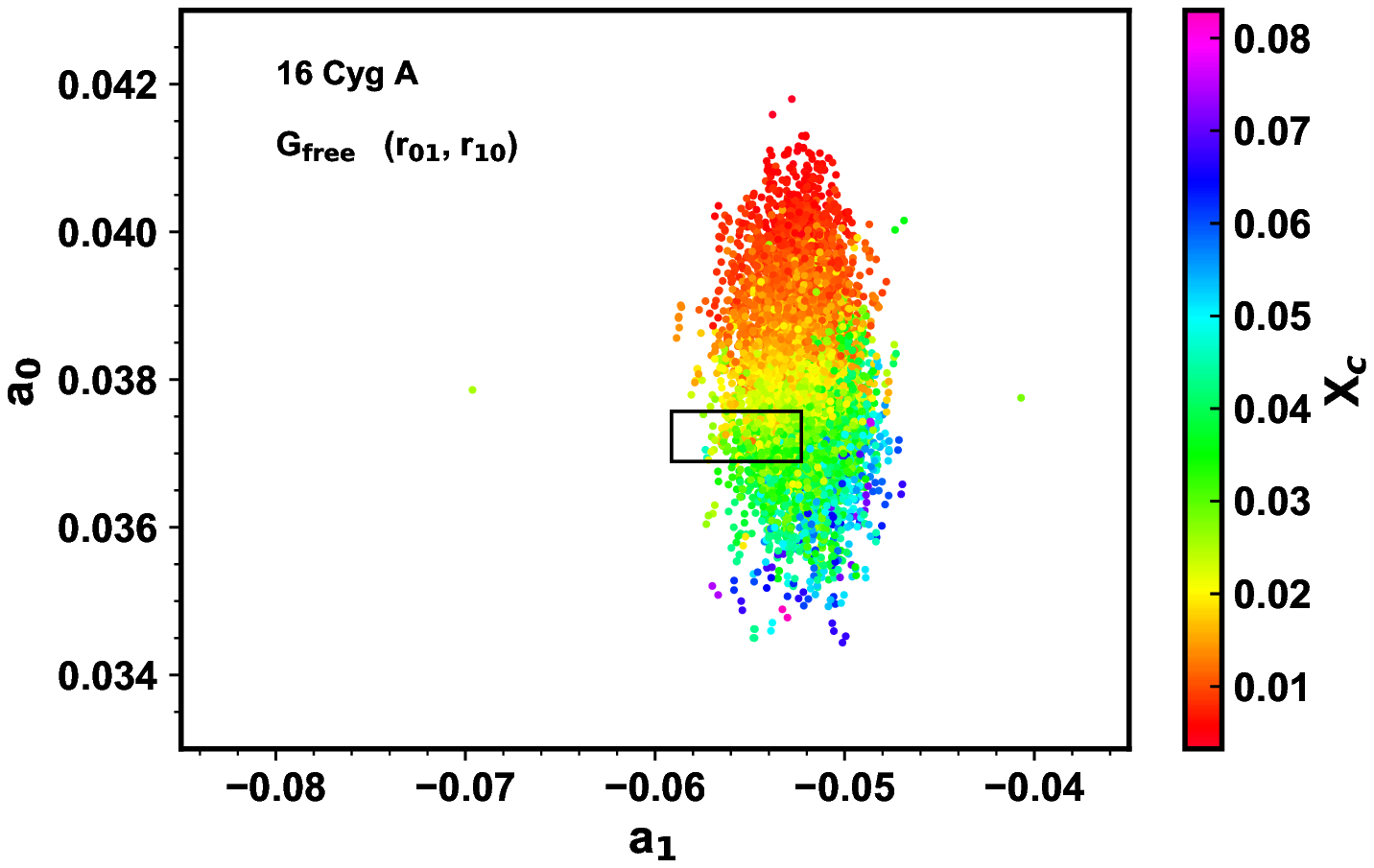}
	\quad
		\includegraphics[width=\columnwidth]{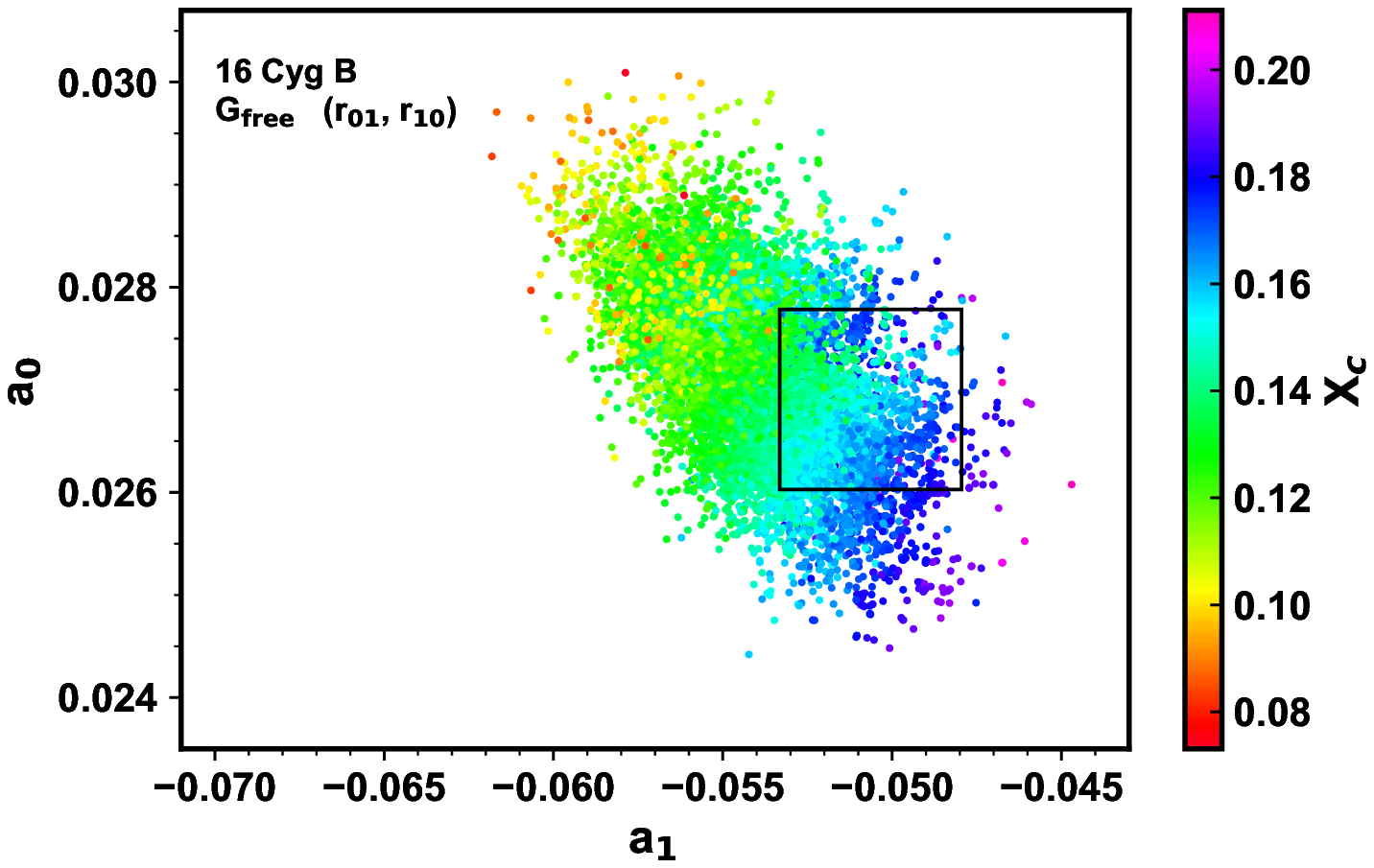}
      \caption{Location of best-fit models for 16 Cyg A (left) and B (right) in the ($a_1$, $a_0$) plane considering both $r_{01}$ and $r_{10}$ fits for grid G$_{1.4}$  (top) and G$_{\rm free}$ (bottom). $a_0$ and $a_1$ are the independent term and linear coefficient of the fitted second order polynomial, respectively. Colours indicate the hydrogen mass fractions in the core. The black box indicates the observational measurements for each star, estimated from $r_{01}$ and $r_{10}$ fits and their $3\sigma$ uncertainties (see text for details).}
   \label{a0a1fits_1}
\end{figure*}
\begin{figure*}
	\includegraphics[width=\columnwidth]{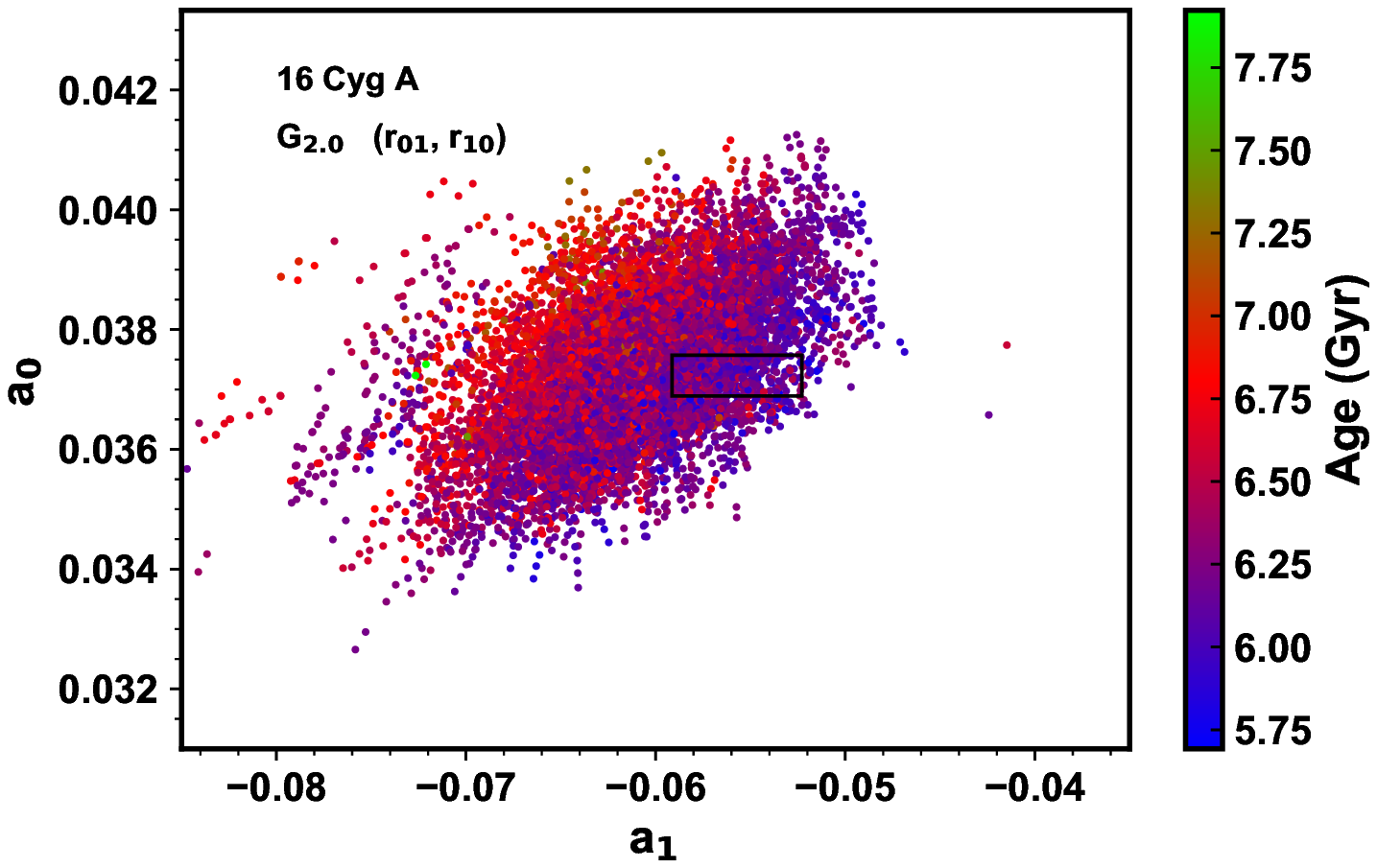}
	\quad
		\includegraphics[width=\columnwidth]{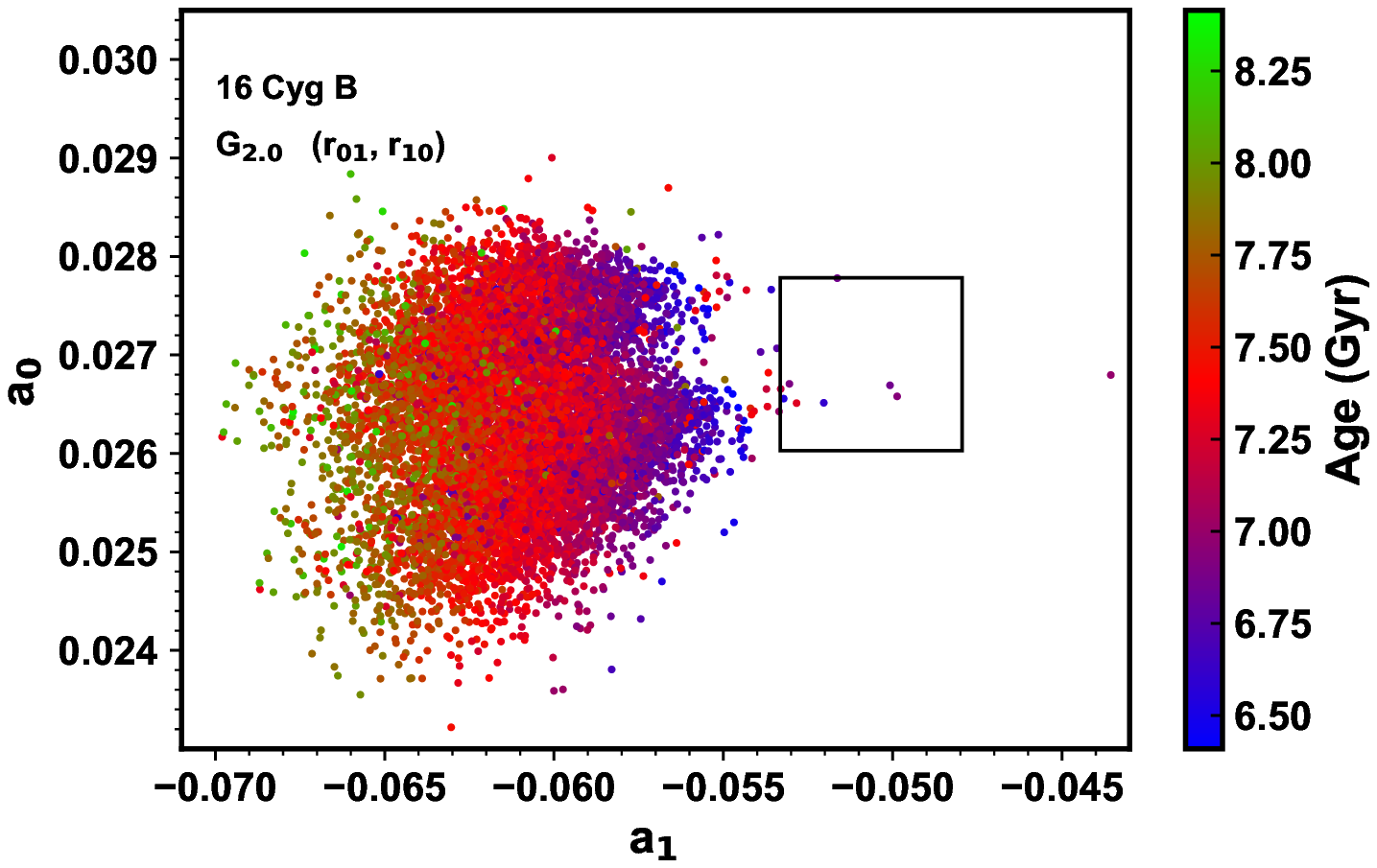}
      \caption{Location of best-fit models for 16 Cyg A (left) and B (right) in the ($a_1$, $a_0$) plane considering both $r_{01}$ and $r_{10}$ fits for grid G$_{2.0}$. $a_0$ and $a_1$ are the independent term and linear coefficient of the fitted second order polynomial, respectively. Colours indicate the stellar ages. The black box indicates the observational measurements 
      for each star, estimated from $r_{01}$ and $r_{10}$ fits and their $3\sigma$ uncertainties (see text for details).}
   \label{fits_1}
\end{figure*}
Figure~\ref{a0a1fits_1} shows the best-fit models of 16 Cyg A and B in the ($a_1$, $a_0$) plane (see Section~\ref{models} for details on how best-fits models were obtained. 
Note that, for each model, there are two estimates of $a_0$ and two estimates of $a_1$, derived from fitting $r_{01}$ and $r_{10}$, respectively. The results are shown for grid G$_{1.4}$ (top panels) and G$_{\rm free}$ (bottom panels), colour-coded according to their corresponding central hydrogen mass fraction, $X_c$. The black box shown in each panel encompasses the two estimates of $a_0$ and $a_1$, including their $3\sigma$ uncertainties.
Similar representation of the best-fit models of 16 Cyg A and B in the ($a_1$, $a_0$) plane for grid G$_{2.0}$ are shown in Figure~\ref{fits_1}, colour-coded according to their corresponding stellar ages.
In all the panels of Figure~\ref{a0a1fits_1}, it can be seen that models tend to clearly separate according to their central hydrogen mass fractions. This is not surprising, since the ratios $r_{01}$ and $r_{10}$ not only carry information about the physical processes near the stellar cores but are also sensitive to the central hydrogen content. Models which occupy the same region as the position of each star (black box symbol) in the ($a_1$, $a_0$) plane shown in Figure~\ref{a0a1fits_1} simultaneously reproduce the observed trend of the ratios and the observational constraints used in the forward modeling process highlighted in Section~\ref{constraints}.

Our findings show that we were unable to find models in grid G$_{2.0}$ which simultaneously satisfy the ratios and other observational constraints used in the forwarding modeling of 16 Cyg B. This is evident in the right panel of Figure~\ref{fits_1}. The top right panel of Figure~\ref{a0a1fits_1} shows that only a few models which simultaneously satisfy the ratios and other observational constraints used in the forwarding modeling of 16 Cyg B were found using grid G$_{1.4}$.
In contrast, we found more  models in grid G$_{\rm free}$ which simultaneously satisfy the ratios ($r_{01}$ and $r_{10}$) and other observational constraints used in the forwarding modeling of 16 Cyg B. 
\begin{figure*}
	\includegraphics[width=\columnwidth]{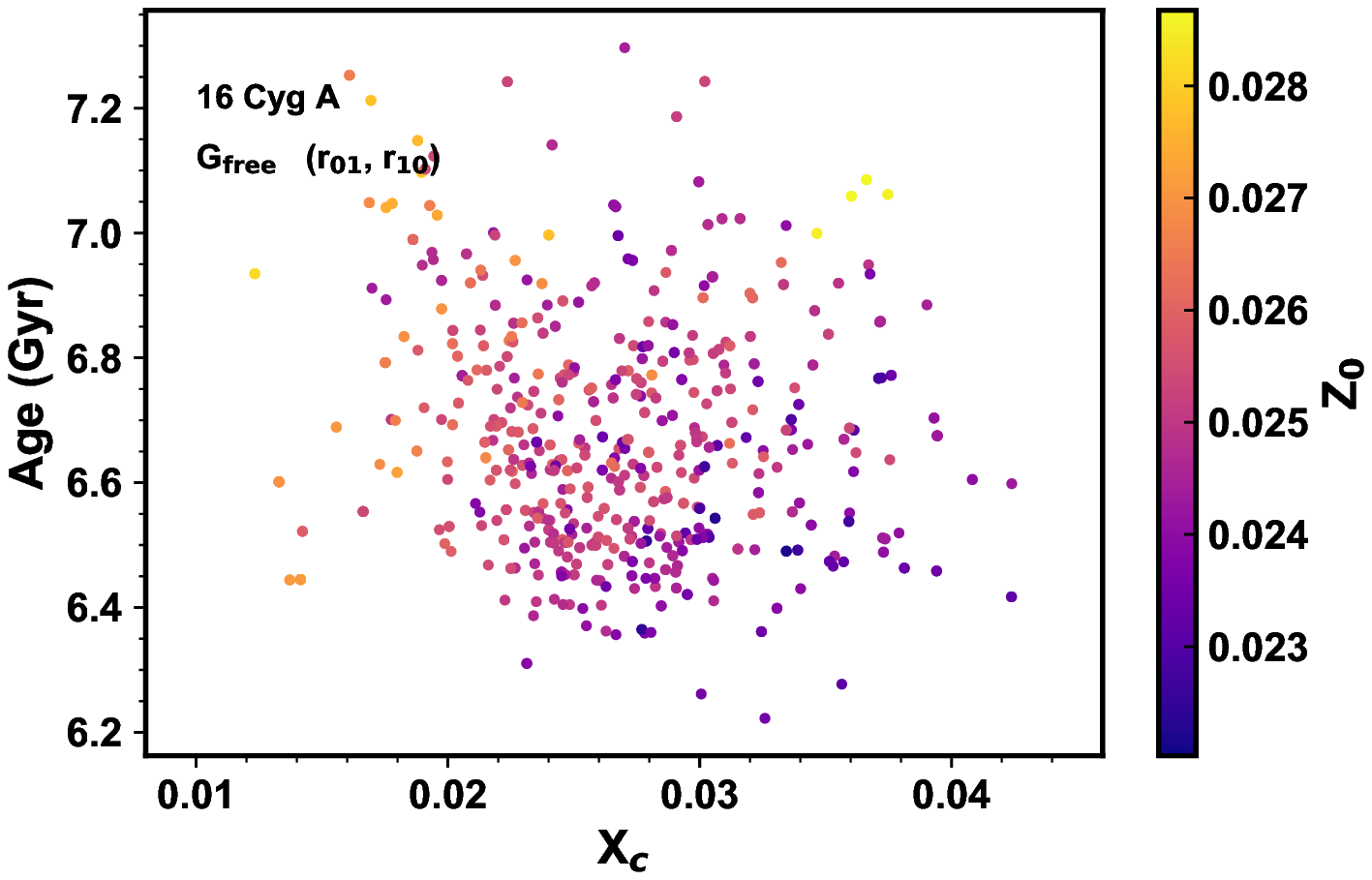}
	\quad
		\includegraphics[width=\columnwidth]{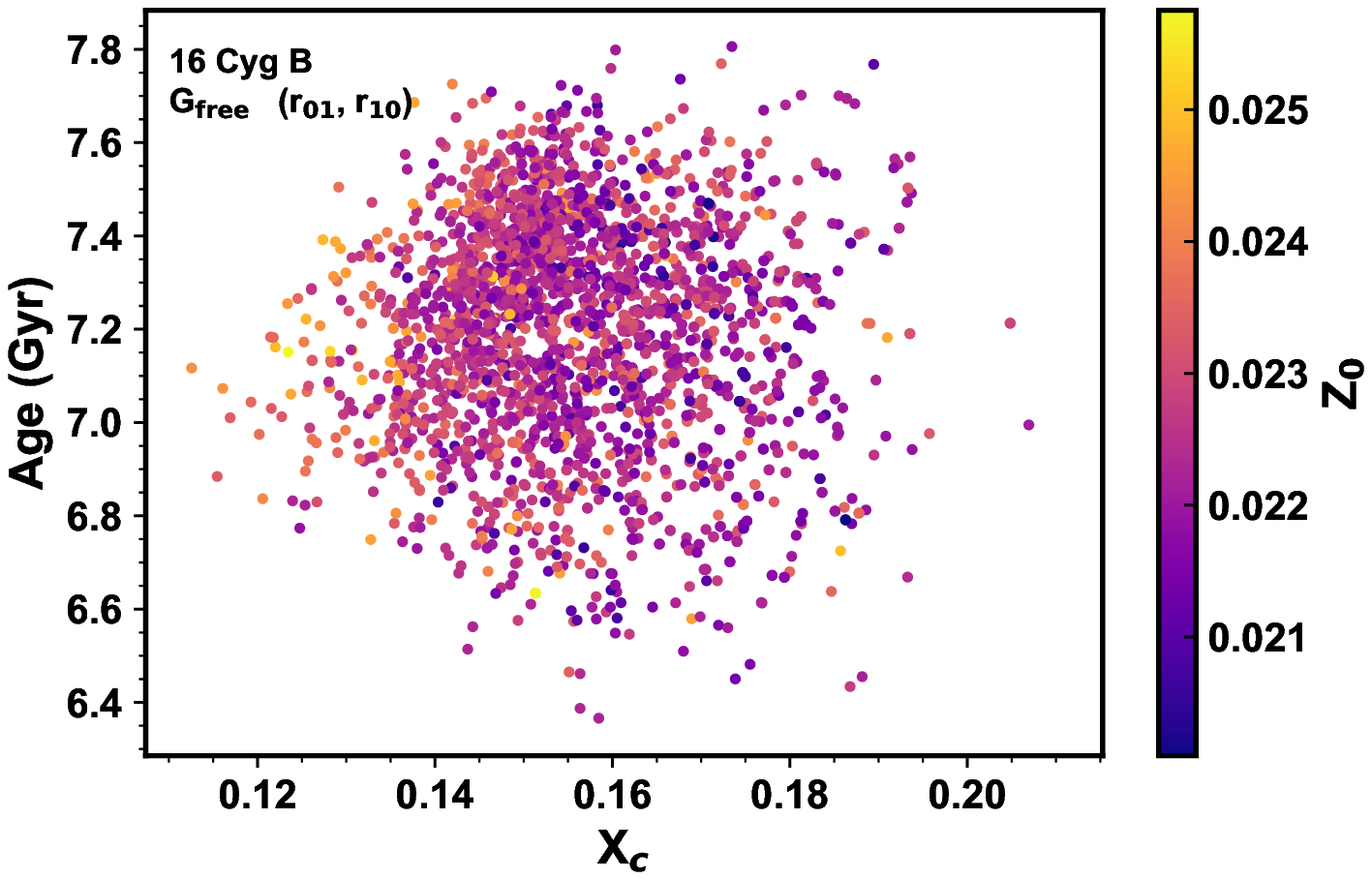}
     \caption{Age vs. X$_c$: Parameters of best-fit models of grid G$_{\rm free}$ which span the 3$\sigma$ uncertainty region of the observed ($a_1$, $a_0$) values determined from $r_{01}$ and $r_{10}$ for 16 Cyg A (left panel) and B (right panel), color-coded according to their corresponding initial metal mass fractions.}
    \label{t_X_AB_1}
\end{figure*}
\begin{figure*}
	\includegraphics[width=\columnwidth]{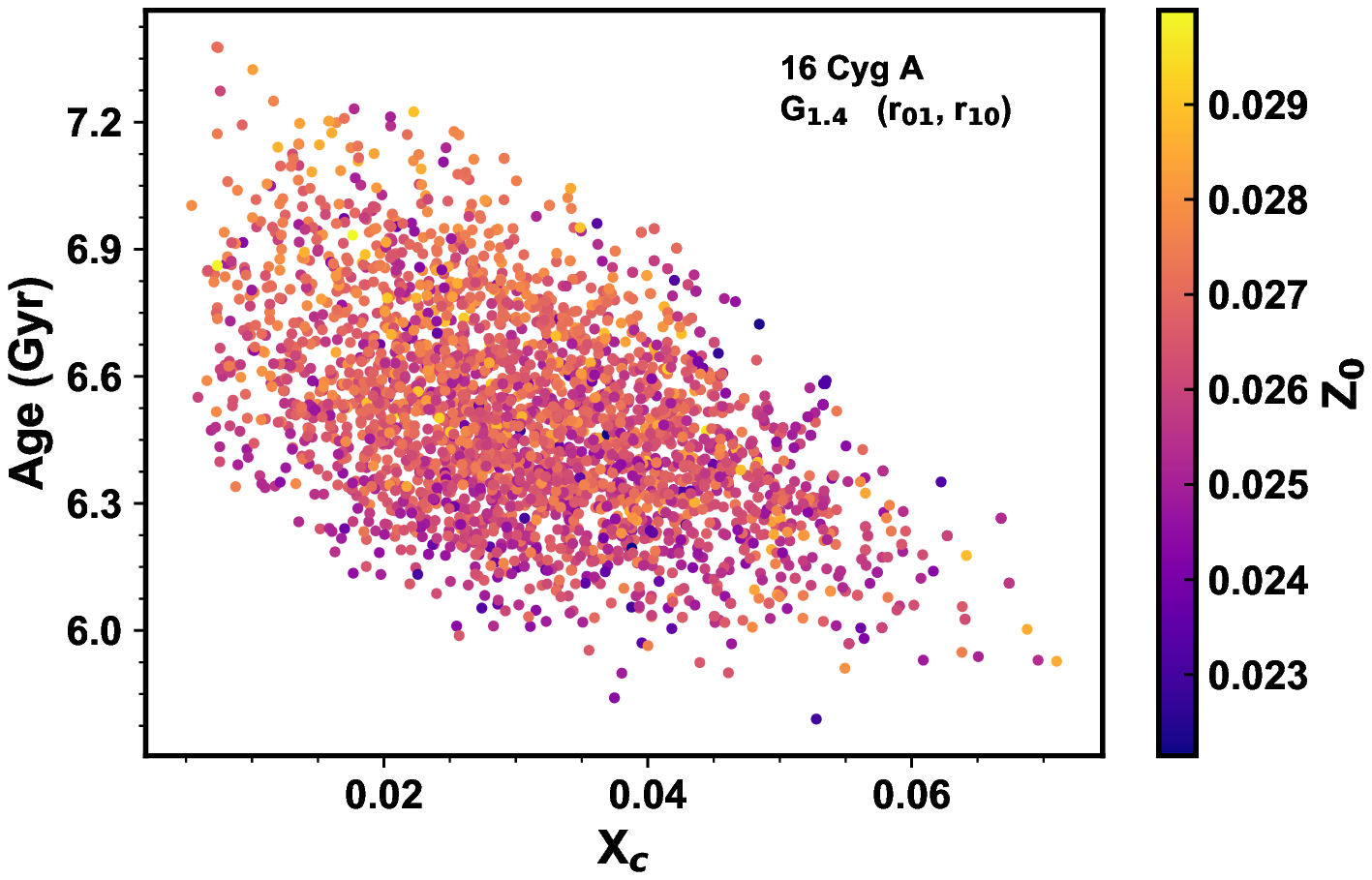}
	\quad
		\includegraphics[width=\columnwidth]{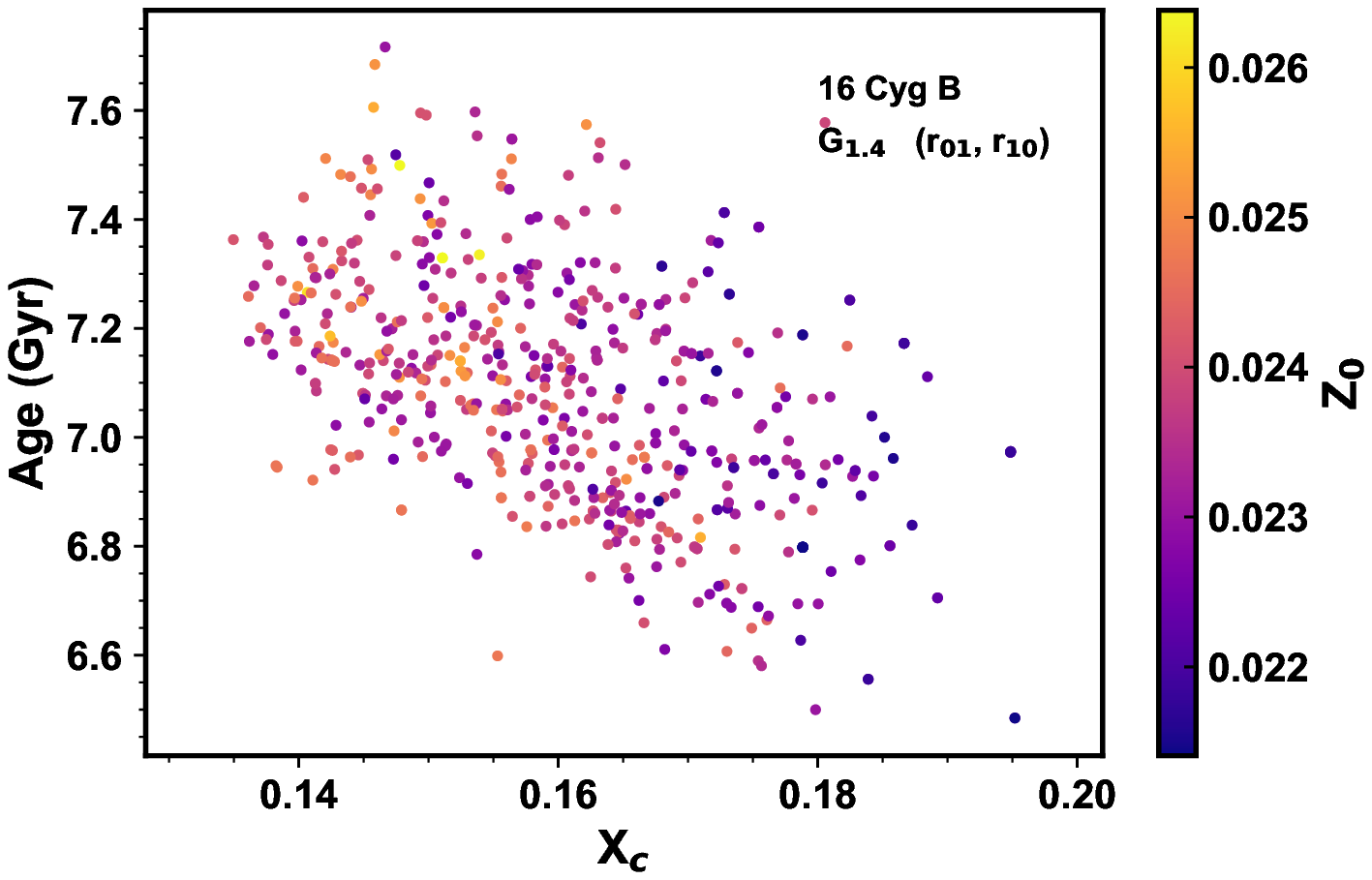}
     \caption{Age vs. X$_c$: Parameters of best-fit models of grid G$_{1.4}$  which span the 3$\sigma$ uncertainty region of the observed ($a_1$, $a_0$) values determined from $r_{10}$ for 16 Cyg A (left panel) and B (right panel), color-coded according to their corresponding initial metal mass fractions.}
    \label{t_X_AB_2}
\end{figure*}
This is shown in the bottom right panel of Figure~\ref{a0a1fits_1}. We stress here that grid G$_{\rm free}$ did not include any restriction in the estimation of the model initial helium mass fraction, while the model initial helium mass fraction in  grids  G$_{1.4}$ and G$_{2.0}$ was determined via a helium-to-heavy element enrichment ratio. 
Therefore, our results further demonstrate that the helium-to-heavy element enrichment law may not be suitable for studies of individual stars. This is consistent with literature findings, e.g. \citet{2021Nsamba, 2021Deal}. 

Next, we extract models for 16 Cyg A and B from grid G$_{\rm free}$ which fall within the  observed ($a_1$, $a_0$) error box of the observed data in the ($a_1$, $a_0$) plane. (see Figure~\ref{t_X_AB_1}). 
Figure~\ref{t_X_AB_2} shows the models extracted in a similar way from grid G$_{1.4}$. Negative linear trends between age and central hydrogen content are evident in the left and right panels of Figure~\ref{t_X_AB_2} for 16 Cyg A and B, respectively. 
Models of 16 Cyg A and B with high central hydrogen mass fractions have lower ages while those with low central hydrogen mass fractions have higher ages. This is expected based on the theoretical description of chemical abundance evolution in main-sequence stars (e.g. \citealt{1990Kippe,2000Prialnik,2010Aerts}). It is worth noting that the linear negative trend is found for grid G$_{1.4}$  (Figure~\ref{t_X_AB_2}) while no such trends are observed in the grid G$_{\rm free}$ (Figure~\ref{t_X_AB_1}). This most probably stems from the linear relation used in grid G$_{1.4}$  to estimate the model abundances via the chemical enrichment ratio (i.e. see Eq.~\ref{law}). Furthermore, this demonstrates that by considering an enrichment law, the impact of stellar aging on the hydrogen abundance is directly visible on the selection of models, while when the initial helium mass fraction,$Y_{i}$, is set to be independent of $Z$, that correlation is diluted through the increase of the possible combinations of stellar chemical content (see Figure~\ref{t_X_AB_1}).

Based on Eq.~(\ref{law}), applying the most optimal initial helium mass fraction values returned from the forward modelling of 16 Cyg A and B using grid G$_{\rm free}$ (i.e. 0.270$\pm$0.013 and 0.263$\pm$0.014, respectively), plus the corresponding optimal initial metal mass fraction values  shown in Table~\ref{grids2} and Table~\ref{grids3}, we deduce that a helium-to-heavy element enrichment ratio of $\sim$0.8 is required to model both components of the 16 Cyg binary system. This value is smaller than the range reported for the Sun by \citet{Serenelli_2010}, i.e. $1.7 \leqslant \Delta Y / \Delta Z  \leqslant 2.2$, depending on the choice of solar composition. However, our inferred value is consistent with the range of values reported by \citet{2019Verma} using a sample of 38 {\it{Kepler}} ``LEGACY'', i.e. $\Delta Y / \Delta Z \in$~[0.38 -- 2.07]. This range was determined  through a combination of surface helium abundances based on the analysis of glitch signatures caused by the ionization of helium, and initial helium abundances determined through abundance differences caused by gravitational settling in stellar models. Furthermore, helium-to-heavy element enrichment ratio values which are below 1 have also been reported in \citet{2015Silva,Aguirre_2017,2021Nsamba}.

From the panels of Figure~\ref{t_X_AB_1} and Figure~\ref{t_X_AB_2}, we see that our down selection of the models leads to a central hydrogen mass fraction for 16 Cyg A spanning the range [0.01 -- 0.06] while for 16 Cyg B we find values within  [0.12 -- 0.19] (cf. panels of Figure~\ref{t_X_AB_1} and Figure~\ref{t_X_AB_2}).
Moreover, the ages of 16 Cyg A and B are found to be in the range [6.0 -- 7.4] and [6.4 -- 7.8], respectively. Finally, the initial metal mass fractions of 16 Cyg A and B lie in the range [0.023 -- 0.029] and [0.021 -- 0.026], respectively.  
Lastly, it can be seen that models which conform to the binary hypothesis that 16 Cyg A and B were born from the same molecular cloud (i.e., implying same initial chemical composition) at approximately the same time, have ages within [6.4 -- 7.4] Gyr and initial metal mass fraction within [0.023 -- 0.026].
We note that these values are based on results emanating from our referencing grid, i.e. grid G$_{\rm free}$ (see panels of Figure~\ref{t_X_AB_1}).


\section{Summary and conclusions}
\label{conclusions}
In this article, we adopted both 16 Cyg A and B as our benchmark stars and presented a novel approach which allows for a selection of a sample of best-fit stellar models returned from forward modelling techniques (involving fitting the observed individual oscillation frequencies and spectroscopic constraints, i.e. metallicity and effective temperature), down to the ones that better represent the core of each star. Our investigations involve using the ratios $r_{01}$ and $r_{10}$ to constrain the central hydrogen mass fraction of 16 Cyg A and B. This is attained by fitting a second order polynomial to the ratios ($r_{01}$ and $r_{10}$) of both models obtained from forward modelling and the observed data of 16 Cyg A and B. By considering the linear ($a_{1}$) and independent ($a_{0}$) coefficients of the second order polynomial fits, we find the models to spread out in the ($a_{1}$, $a_{0}$) plane according to their respective central hydrogen mass fractions.

This approach allowed for a selection of models which simultaneously satisfy the ratios ($r_{01}$ and $r_{10}$) and other observational constraints used in the forwarding modelling process. This implies that these selected models also satisfy the interior conditions of our target stars. Following this approach, we find that that the central hydrogen content of 16 Cyg A and B lie in the range [0.01 -- 0.06] and [0.12 -- 0.19], respectively. Moreover, a common age and initial metal mass fraction for the two stars requires these parameters to lie in the range [6.4 -- 7.4] Gyr and [0.023 -- 0.026], respectively.

Furthermore, our findings show that grids having the model initial helium mass fraction determined via a helium-to-heavy element enrichment ratio may not always include models that satisfy the core conditions set by the observed ratios. This was particularly evident when modelling 16 Cyg B.




\section*{Acknowledgements}

The authors acknowledge the dedicated team behind the NASA'S {\it{Kepler}} missions. BN thanks Achim Weiss, Earl P. Bellinger, Selma E. de Mink, and
the Stellar Evolution research group at Max-Planck-Institut f\"{u}r Astrophysik (MPA) for the useful comments on this article.
B.N. also acknowledges postdoctoral funding from the Alexander von Humboldt Foundation and "Branco Weiss fellowship -- Science in Society" through the SEISMIC stellar interior physics group.
M. S. Cunha is supported by national funds through Funda\c{c}\~{a}o para a Ci\^{e}ncia e a Tecnologia (FCT, Portugal) - in the form of a work
contract and through the research grants UIDB/04434/2020, UIDP/04434/2020 and PTDC/FIS-AST/30389/2017,
and by FEDER - Fundo Europeu de Desenvolvimento Regional through COMPETE2020 - Programa Operacional Competitividade e Internacionaliz\'{a}cao (grant: POCI-01-0145-FEDER-030389).
T.L.C.~is supported by Funda\c{c}\~{a}o para a Ci\^{e}ncia e a Tecnologia (FCT) in the form of a work contract (CEECIND/00476/2018).
We thank the reviewer for the constructive remarks.

\section*{Data Availability}
 
Data used in this article are available in machine-readable form.
The details of the ``inlist'' files used in the stellar evolution code (MESA) are described in \citet{2021Nsamba}.



\bibliographystyle{mnras}
\bibliography{mnras} 




\appendix
\section{Polynomial coefficients of the best-fit models}
\label{model_a1a0}
$a_{1}$ and $a_{0}$ values for the best-fit models of grid G$_{1.4}$, G$_{2.0}$, and G$_{\rm free}$ are available in the machine-readable form in the online version of this paper.


\bsp	
\label{lastpage}
\end{document}